\documentclass[10pt]{elsart}
\usepackage[english]{babel}
\usepackage{latexsym}
\usepackage{rotating}
\usepackage{amsmath}
\usepackage{amssymb}
\usepackage{lscape}
\usepackage{epsfig}
\usepackage{color}
\usepackage{natbib}
\usepackage[latin1]{inputenc}
\newcommand{\be}{\begin{eqnarray*}}
\newcommand{\ee}{\end{eqnarray*}}
\newcommand{\ben}{\begin{eqnarray}}
\newcommand{\een}{\end{eqnarray}}
\newcommand{\uE}[1]{\mathbb{E}\left[ #1 \right] } 

\begin{document}

  \begin{frontmatter}

    \title{{Bayesian Value-at-Risk with Product Partition Models \thanksref{cor1}}}

    \author[1,2]{Giacomo Bormetti },
    \author[3]{Maria Elena De Giuli},
    \author[4,2]{Danilo Delpini},
    \author[3]{Claudia Tarantola}

    \thanks[cor1]{Address for correspondence: Giacomo Bormetti, National Institute of Nuclear Physics - Pavia Unit,
      Via A. Bassi 6, 27100 Pavia, Italy, \mbox{Giacomo.Bormetti@pv.infn.it}, tel: 0382 987703, fax: 0382 526938 .}

    \address[1]{Institute for Advanced Studies - Center for Risk and Security Study\\
    Viale Lungo Ticino Sforza 56, 27100 Pavia, Italy}
    \address[2]{National Institute of Nuclear Physics - Pavia Unit\\
    Via A.Bassi 6, 27100 Pavia, Italy}
    \address[3]{Department of Economics and Quantitative Methods, University of Pavia\\
    Via San Felice 5, 27100 Pavia, Italy}
    \address[4]{Department of Nuclear and Theoretical Physics, University of Pavia\\
    Via A.Bassi 6, 27100 Pavia, Italy}

    \date{}

    \begin{abstract}
      In this paper we propose a novel Bayesian methodology for Value-at-Risk computation based on parametric
      Product Partition Models. Value-at-Risk is a standard tool to measure and control the
      market risk of an asset or a portfolio, and it is also required for regulatory purposes.
      Its popularity is partly  due to the fact that it is an easily understood
      measure of risk.
      The use of Product Partition Models allows us to  remain in a
      Normal setting even in presence of outlying points,
      and to obtain a closed-form expression for Value-at-Risk
      computation. We present and compare two different
      scenarios:  a product partition structure on
      the  vector of means and a product partition structure  on the vector of variances.
      We apply our methodology to an Italian stock
      market data set from Mib30. The numerical results clearly show that
      Product Partition Models can be successfully exploited in order to quantify market risk
      exposure. The obtained Value-at-Risk estimates are in full agreement with Maximum Likelihood
      approaches, but our methodology provides richer information about the clustering structure
      of the data and the presence of outlying points.

    \end{abstract}


  \end{frontmatter}

  \section{Introduction}

  Following the increase in financial uncertainty, there has been
  intensive research from financial institutions, regulators and
  academics to  develop  models  for  market risk evaluation. A common and 
  easily understood measure of risk is Value-at-Risk (VaR).
  In particular, Basel accords impose that all financial institutions
  have to meet capital requirements based on VaR estimates, see
  \cite{Basel:2006}.

  VaR is defined as the maximum potential loss of an asset or a
  portfolio, at a given time horizon and  significance level.
  An  accurate  estimate of VaR is important for both banks and
  regulators.
  An underestimation of risk could obviously cause
  problems for banks and other participants in financial markets
  (e.g. bankruptcy). On the other hand, an overestimation of risk
  may cause one to allocate too much capital as a cushion for risk
  exposures, having a negative effect on profits. The Committee does
  not prescribe banks a special type of model,  leaving them free to
  specify their own model for VaR estimation. In the  literature a
  wide range of models to measure VaR and to determine
  the level of regulatory capital are described. For  a review on VaR models see e.g.
  \cite{Jorion:2001}, \cite{Manganelli:2004} and the remaining list of VaR contributions at the
  web site \emph{http://www.gloriamundi.org}.

  In this paper we propose a novel Bayesian methodology for VaR
  estimation based on parametric Product Partition Models (PPMs), and
  we compare our results with those obtained with standard
  approaches based on Maximum Likelihood (ML) techniques,
  see e.g. \cite{Mina_Xiao:2001}, \cite{Mattedi_etal:2004}, \cite{Bormetti:2007} .
  In our analysis we pay particular attention in evaluating the
  statistical
  uncertainty associated with different results; in fact  a good risk management requires not only
  a pointwise VaR estimate but also an assessment of how much precise
  the estimate is.

  If the returns are    independent and identically  normally
  distributed a closed-form and easy to implement expression
  for VaR can be used.
  Unfortunately, these assumptions fail to be effective for low liquidity markets and
  short time horizons and have to be relaxed. Possible solutions are
  to resort to heavy tailed distributions or to abandon the
  hypothesis of identically distributed returns. In this paper we
  follow the latter approach and we use a Bayesian methodology based
  on parametric PPMs.
  We assume that the returns follow a Normal distribution with a
  partition structure on the parameters of interest. We assign a
  prior distribution on the space of all possible partitions and we
  identify clusters of returns sharing the same mean and variance
  values. Returns belonging to different clusters are characterised
  by different values either of the  mean or the variance. The hypothesis of identical
  distribution holds within but non between clusters. As a consequence
  we abandon the assumption of identical distribution while preserving
  the Normal setting. Furthermore, the use of a product partition approach
  allows us not only to accommodate anomalous observations
  but also provides as a by-product a useful tool for their
  identification; see \cite{Quintana_Iglesias:2003},
  \cite{Quintana_etal:2005} and \cite{DeGiuli_etal:2009} for further
  details.

  We propose and compare two different
  PPMs for VaR estimation. In the first one we impose a partition
  structure on the vector of means whereas the volatility is a common random
  variable;  in the second one we impose a partition structure on the vector of variances with common unknown mean.
  The first approach is quite
  effective for VaR estimation, but it is very sensitive to the
  values of prior parameters and even a hierarchical model can not
  reduce this sensitivity. This problem can be  overcame by fixing
  the values of the hyperparameters according to analysts'
  experience about the market behaviour.
  This drawback effect is strongly reduced by imposing a partition
  structure on the vector of variances. Our results are compared
  with those obtained with the  parametric PPM developed in
  \cite{Loschi_etal:2003} for the identification of change-points in financial time
  series.
  
  To obtain the posterior distribution of the quantity of
  interest we use Markov Chain Monte Carlo (MCMC) techniques.
  MCMC methods have a history in mathematical physics dating back to the algorithm
  of \cite{Metropolis_etal:1953}, later generalised by \cite{Hastings:1970}.
  In our work we extensively resort to a specific type of Markov chain algorithm,
  introduced by \cite{Geman:1984} and \cite{Gelfand:1990} and known as Gibbs
  sampler. The Gibbs sampling algorithms considered here are
  described in details in sections \ref{s:means} and
  \ref{s:variances}. For a recent Bayesian application of Gibbs sampling in the
  context of financial analysis, see e.g. \cite{Chang_etal:2008}.

  The paper is organized as follows. In section \ref{s:back} we
  briefly introduce  VaR as a measure of risk  and parametric PPMs. In section
  \ref{s:PPM} we present two models for VaR estimate and introduce a closed-form
  expression for VaR computation extending the usual Gaussian form. In section \ref{s:Out} we describe
  how to  exploit the clustering structure induced by PPMs in order
  to identify outlying points. In section \ref{s:numerical} we apply
  our methodologies to a Mib30 data set and provide a sensitivity
  analysis of our results with respect to different choices of
  hyperparameters. Section \ref{s:conclusions} closes the paper with
  some final  remarks.

  \section{Background and Preliminaries}\label{s:back}

  \subsection{Value-at-Risk}

  VaR is referred  to the probability of extreme losses
  due to adverse market movements.
  In particular,
  for a given significance level ${\alpha}$ (typically $1\%$ or $5\%$),
  VaR is defined as the maximum potential loss over a fixed time horizon
  of individual assets and portfolios of assets as well.
  In the following we focus  on VaR for a single asset.

  If the returns are independent and identically normally distributed with mean $\mu$ and
  variance $\sigma^2$, a closed-form expression for VaR normalised to the spot price is
  given by
  \ben\label{eq:normalVaR}
  \frac{\Lambda}{W_0}=-\mu+\sigma \sqrt{2} \;\mathrm{erfc}^{-1}\left(2
  \alpha\right),
  \een

  where $\Lambda$ is VaR, $W_0$ is the spot price and
  $\mathrm{erfc}^{-1}$ is the inverse of the complementary error function.
  In the following, with VaR we shall refer to the quantity $\Lambda/W_0$,
  if not specified otherwise. 
  If this quantity is expressed in percentage term we name it percentage VaR,
  \mbox{VaR$(\%)$} .

  In order to estimate the parameters $\mu$ and $\sigma$ in equation
  (\ref{eq:normalVaR}), we apply a Bayesian approach based on
  parametric PPMs; the details are provided in the following section.

  \subsection{Parametric Product Partition Models}

  We now briefly review the theory of parametric PPMs  with reference to our specific problem.
  For a detailed and more general presentation see \cite{Hartigan:1990}, \cite{Barry_Hartigan:1992}.

  Let $\boldsymbol{y}=\left(y_{1},
  \ldots,y_{t},\ldots y_{T}\right)$ denote
  the vector of returns of a generic
  asset at different time points $t$. The returns are independent, and jointly distributed with  probability density function $f$ parameterised
  by the vector
  $(\boldsymbol{\theta},{\psi})$. The elements of
  $\boldsymbol{\theta}$ depend on the time point $t$,
  $\boldsymbol{\theta}=(\theta_{1},\ldots,\theta_{T}) $,
  whereas   ${\psi}$ is a   parameter  that is common to all observations.
  We consider the following model

 \begin{equation}
\boldsymbol{y}|(\boldsymbol{\theta},\psi) \thicksim
f(\boldsymbol{y}|\boldsymbol{\theta},\psi), \;\;
\;\;{\mbox{with}}\;\; \;\;y_t \stackrel{ind}{\thicksim}
f(y_t|\theta_t, {\psi} ) \; \; \; \; \;
  t=1,\ldots,T. \label{cla1}
  \end{equation}

  Given the model in (\ref{cla1}), let  $S_{0}=\{{t}:
  t=1,\ldots, {T} \}$ be the set of all time periods.
  A partition of the  set $S_0$,  $\rho = \left\{ S_{1}, \ldots,
  S_{d}, \ldots, S_{|\rho|} \right\}$ with cardinality $|\rho|$,
  is defined by the property that $S_d \cap
  S_{d^{\prime}}=\emptyset$ for $d \neq d^{\prime}$ and $\cup_d\,
  S_d=S_0$. The generic element of $\rho$ is $S_{d}=\left\{ t:
  \theta_{t}=\theta^{*}_{d}\right\}$, where
  \mbox{$\boldsymbol{\theta}^*=\left(\theta^{*}_{1},\ldots,\theta^{*}_{|\rho|}\right)$}
  is the vector of the unique  values of $\boldsymbol{\theta}=\left(\theta_{1},\ldots,\theta_{T}\right) $.
  All $\theta_t$ whose subscripts $t$ belong to the same set $S_d \in \rho$ are (\emph{stochastically})
  equal, in this sense they are regarded as a single \emph{cluster}.

  We assign to each partition $\rho$ the following  prior
  distribution
  \ben
  P\left(\rho=\left\{ S_{1}, \ldots,S_{|\rho|}
  \right\}\right)=K \prod_{d=1}^{|\rho|} C\left(S_{d}\right)~,
  \label{eq:product_distribution_rho}
  \een
  where $C\left(S_{d}\right)$ is a \emph{cohesion function} and $K$
  is the normalising constant. Equation
  (\ref{eq:product_distribution_rho}) is referred to as the
  \emph{product distribution} for partitions. The cohesions
  represent prior weights on group formation and  formalise our
  opinion on how tightly clustered the elements of $S_{d}$ would be.

  The cohesions can be specified in different ways. A useful choice is
  \ben
  \label{eq:cohesion}
  C\left(S_d\right)= c \times
  \left(\left|S_d\right|-1\right)!~,
  \een
  where  $c$ is a positive constant and $\left|S_d\right|$ denotes the cardinality of the set $S_d$. For moderate values of $c$,
  e.g.\ $c=1$, the cohesions in equation~(\ref{eq:cohesion}) yield a prior
  distribution that favours the formation of partitions with a
  reduced number of large subsets. For more details on the choice of
  $c$ see e.g. \cite{Liu:1996},
  \cite{Quintana_Iglesias:2003}, \cite{Quintana_etal:2005} and
  \cite{Tarantola_etal:2008}.

  If non contiguous clusters are considered we can exploit  an interesting connection between parametric
  PPMs and the class of Bayesian nonpara\-me\-tric models with
  a Dirichlet Process prior, see \cite{Antoniak:1974}.
  Under the latter prior, the marginal distribution of the observables is
  a specific PPM with the cohesion functions specified by
  equation~(\ref{eq:cohesion}), see \cite{Quintana_Iglesias:2003}.
  In this case we can use efficient Markov Chain Monte Carlo (MCMC) algorithms
  developed for Bayesian nonparametric problems.

  When dealing with contiguous blocks, as in the change-point problem, this connection cannot be
  exploited, and specific MCMC algorithms are required,
  see e.g. \cite{Loschi_etal:2003}.

  \section{VaR Computation via Product Partition Models}\label{s:PPM}

  Let $\boldsymbol{y}$ be the vector of daily returns of a generic
  asset. We assume that the returns are normally distributed with
  parameter vector $(\boldsymbol{\theta},{\psi})$. We present and
  compare  two different PPMs; in the first one we impose  a
  partition structure on the  vector of means, and in the second one
  we  consider partitions on the vector of variances. In the
  following the  PPM applied to the vector of means will be shortly
  referred to as the $\boldsymbol{\mu}$-PPM approach, while
  $\boldsymbol{\sigma^2}$-PPM will refer to the PPM for  the vector
  of variances.
  In  $\boldsymbol{\mu}$-PPM  the vector $\boldsymbol{\theta}$ is
  the vector of means while in $\boldsymbol{\sigma^2}$-PPM it
  corresponds to the  vector of variances. In the former
  model ${\psi}$ is the variance and in the latter it corresponds to
  the mean.

  We consider the following hierarchical  structure
  \begin{eqnarray*}
    &\;&y_t| (\rho,(\theta^*_{{1}},\dots,\theta^*_{{|\rho|}}),
    \sigma^2) \stackrel{ind.}{\thicksim} N(y_t|(\theta _{t},\psi))~,\\
    &\;& \theta^*_{{1}},\dots,\theta^*_{{|\rho|}}|(\rho,\psi)
    \stackrel{i.i.d.}{\thicksim}f(\cdot|\psi)~,\\
    &\;&  \rho \sim \mbox{product distribution, with}\; \;
    C(S_d)=c \times (|S_d|-1)!~,\\
    &\;& \psi\thicksim g(\psi)~,
  \end{eqnarray*}
  where $f$ and $g$ denote generic density functions and the
  product distribution is defined in
  equation~(\ref{eq:product_distribution_rho}).

  The elicitation of a partition structure on the  vector of means (or variances)
  allows us to remain in a Normal setting without assuming identical
  distribution of the returns. An alternative model that could be used
  to take into account atypical returns when estimating VaR consists of assuming
  $t$-distributed rather than Normal data. This analysis has been performed by
  \cite{Quintana_Iglesias:2003} in the context of regression models,
  showing that parametric PPMs in a Normal setting are even more effective
  if the purpose is to deal and identify   extreme values.

  In sections \ref{s:means} and
  \ref{s:variances} we describe in details our models and in section
  \ref{s:VAR} we propose a closed-form expression for VaR
  computation.

  \subsection{Product Partition Models on Vector of Means}\label{s:means}

  In the $\boldsymbol{\mu}$-PPM approach we impose a partition
  structure on the  vector of means
  $\boldsymbol{\mu}=(\mu_1,\ldots,\mu_T) $. By inducing a cluster
  structure on the vector $\boldsymbol{\mu}$ we try to
  accommodate for atypical $y_t$ values.
  In order to achieve this goal we use the following hierarchical model
  \begin{eqnarray}
    \label{mod:mean}
    &\;&y_t| (\rho,(\mu_1^{*},\dots,\mu_{|\rho|}^{*}),
    \sigma^2) \stackrel{ind.}{\thicksim} N(\mu _{t},\sigma ^{2})~,\nonumber\\
    &\;& \mu_1^{*},\dots,\mu_{|\rho|}^{*}|(\rho,\sigma^{2})
    \stackrel{i.i.d.}{\thicksim} N(m,\tau _{0}^{2}\sigma ^{2})~,  \\
    &\;&  \rho \sim \mbox{product distribution, with}\; \;
    C(S_d)=c \times (|S_d|-1)!~, \nonumber \\
    &\;& \sigma ^{2}\thicksim IG(\nu_{0},\lambda_{0})~,\nonumber
  \end{eqnarray}
  where $\boldsymbol{\mu}^*=(\mu^*_{{1}},\dots,\mu^*_{|\rho|}) $ is the
  vector of all entries of $\boldsymbol{\mu}$ for a
  given partition $\rho$, and $IG(\nu_{0},\lambda_{0})$ is an
  Inverted Gamma distribution with
  $\uE{\sigma^2}=\lambda_0/(\nu_0-1)$, $\nu_0>1$ and $\lambda_0>0$.

  The  complete joint distribution for the model  is given by
  \ben
  &&f(\boldsymbol{y},\boldsymbol{\mu},\rho=
  \{S_{1},\dots,S_{|\rho|}\},\sigma^2)\ \propto\ \exp\left\{
  -\frac{1}{2\sigma^2}\sum_{d=1}^{|\rho|}\sum_{t\in S_{d}}\left(y_t-\mu^*_{{d}}\right)^2
  \right\}\nonumber\\
  &&\;\;\;\times\exp\left\{
  -\frac{1}{2\tau_0^2\sigma^2}\sum_{d=1}^{|\rho|}(\mu^*_{{d}}-m)^2-\frac{\lambda_0}{\sigma^2}
  \right\}\frac{1}{(\sigma^2)^{1+\nu_0+\frac{T+\rho}{2}}}\prod_{d=1}^{|\rho|}
  (|S_{d}|-1)!~.\nonumber
  \een
  To fit this model we adapt an algorithm proposed by
  \cite{Bush_MacEachern:1996} in the context of Bayesian
  nonparametric inference.
  Once a starting value for the vector
  $\boldsymbol{\mu}$ has been provided, we iteratively sample from
  the joint posterior distribution of model and parameters by means
  of the Gibbs algorithm described below.
  \begin{itemize}
  \item[\emph{Step} (i)~:]
    Sample $\sigma^2$ from its full conditional distribution
    $$\sigma^2|\boldsymbol{\mu},\boldsymbol{y}\sim IG\left\{\nu_0+\frac{T}{2}+\frac{|\rho|}{2},\lambda_0+
    \frac{1}{2\tau_0^2}\sum^{|\rho|}_{d=1}(\mu_{d}-m)^2+\frac{1}{2}\sum_{t=1}^T(y_{t}-\mu_{t})^2\right\}.
    $$
  \item[\emph{Step} (ii)~:] Update each  $\mu_{t}$, $t=1,\ldots,T$,
    by sampling  from the mixture
    \ben
    &&\mu_{t}|\boldsymbol{\mu}_{-t},\sigma^2, \boldsymbol{y}\sim\sum_{j\neq t} q_{tj}
    \delta_{\mu_{j}}(\mu_{t})+q_{t0} \times N\hspace{-0.1cm}\left(\frac{y_{t}\tau_0^2+m}
    {1+\tau_0^2},\frac{\sigma^2\tau_0^2}{1+\tau_0^2}\right),
    \label{eq:upmean}
    \een
    where $\boldsymbol{\mu}_{-t}$ is obtained from  $\boldsymbol{\mu}$ by removing the  $t$-th entry and
    $\delta_{\mu_j}(\mu_t)$ is the Dirac delta centered on $\mu_t$.

    \noindent The  distribution  in equation~(\ref{eq:upmean}) corresponds to a mixture of point
    masses and a Normal distribution, with weights
    \begin{eqnarray*}
      &&q_{tj}\propto \exp
      \left \{{-\frac{1}{2\sigma^2}(y_{t}-\mu_{j})^2}\right\},\\
      &&
      q_{t0}\propto\frac{c}{\sqrt{1+\tau_0^2}}\exp\left\{{-(y_{t}-m)^2/[2\sigma^2(1+\tau_0^2)]}\right\},
      \\
      &&\sum_{j\neq t} q_{tj}+q_{t0}=1.
    \end{eqnarray*}
  \item[\emph{Step} (iii)~:] Before proceeding to the next Gibbs iteration we update the vector
    $\boldsymbol{\mu}^*$, given the partition $\rho$, sampling from
    $$
    \mu_{d}^*\sim N \hspace{-0.1cm}\left(\frac{\sum_{t \in S_d}y_{t}+m/\tau_0^2}
    {|S_d|+1/\tau_0^2},\frac{\sigma^2}{|S_d|+1/\tau_0^2}\right)\;\;\;d=1,\ldots,|\rho|.
   $$
    This last step was introduced in \cite{Bush_MacEachern:1996} to
    avoid being trapped in sticky patches in the Markov space.
  \end{itemize}

  The weights $q_{tj}$ represent the finite probability of replacing
  $\mu_t$ with a value $\mu_j$ already belonging to the vector
  of means. On the other hand $q_{t0}$ represents  the finite
  probability of replacing the old $\mu_t$ value with a newly sampled
  one. It is worth noticing again the role played by the constant $c$. A greater value of $c$ increases the probability to
  generate new values. Generally, the higher   $c$ is, the higher
  the probability to obtain an elevate number of clusters will
  be.

  As it turns out from the empirical analysis, see section
  \ref{s:numerical}, posterior distributions are quite sensitive to
  the value of the parameter $\lambda_0$ of the Inverted Gamma
  distribution in the hierarchical model defined in~(\ref{mod:mean}).
  We tried to reduce this
  drawback effect by introducing a hyperprior distribution on
  $\lambda_0$. This translates into a minor modification of model
  (\ref{mod:mean})
  \begin{eqnarray}
    \label{mod:mean-ger}
    &\;&y_t| (\rho,(\mu_1^{*},\dots,\mu_{|\rho|}^{*}),
    \sigma^2,\lambda_0) \stackrel{ind.}{\thicksim} N(\mu _{t},\sigma ^{2})~,\nonumber \\
    &\;& \mu^*_{{1}},\dots,\mu^*_{{|\rho|}}|(\rho,\sigma^{2},\lambda_0)
    \stackrel{i.i.d.}{\thicksim} N(m,\tau _{0}^{2}\sigma ^{2})~,\\
    &\;&  \rho \sim \mbox{product distribution}~,\nonumber\\
    &\;& \sigma ^{2}|\lambda_0\thicksim IG(\nu_{0},\lambda_0)~, \nonumber\\
    &\;& \lambda_0\thicksim G(\eta,\phi)~, \nonumber
  \end{eqnarray}
  where $G(\eta,\phi)$ is
  a Gamma distribution with $\uE{\lambda_0}=\eta\phi$,
  $\eta>0$ and $\phi>0$. The previous Gibbs sampling algorithm must
  be modified coherently. Now we have to provide a starting point
  for $\sigma^2$ too, while \emph{Step} (i) splits in two
  sub-steps:
  \begin{itemize}
  \item[\emph{Step} (i\rm{a})~:]
    \begin{eqnarray}
      \label{eq:hyperlambdafullcond}
      \lambda_0|\boldsymbol{\mu},\boldsymbol{y},\sigma^2\sim
      G\left(\nu_0+\eta,\frac{\sigma^2\phi}{\sigma^2+\phi}\right).
    \end{eqnarray}
  \item[\emph{Step} (i\rm{b})~:]
    \begin{eqnarray*}
      \sigma ^{2}| \boldsymbol{\mu},\boldsymbol{y},\lambda_0\sim IG\left(\nu
      _{0}+\frac{|\rho|}{2}+\frac{T}{2}, \lambda_0+\frac{1}{2\tau
    _{0}^{2}}\sum_{d=1}^{|\rho|}(\mu_{{d}}-m)^{2}+
      \frac{1}{2}\sum_{t=1}^{T}(y_{t}-\mu _{t})^{2}\right).%\nonumber\\
    \end{eqnarray*}
  \end{itemize}
  \emph{Step} (ii) and \emph{Step} (iii) do not change.

  \subsection{Product Partition Models on Vector of Variances}\label{s:variances}

  An alternative way to relax the hypothesis of identical
  distribution of the returns, without renouncing to the normality
  assumption, is to promote the variance $\sigma^2$ from a scalar to
  the vectorial quantity
  $\boldsymbol{\sigma^2}=(\sigma^2_1,\ldots,\sigma^2_T) $ and to
  impose a clustering structure on $\boldsymbol{\sigma^2}$. Our aim is to
  create clusters of observations, not necessarily contiguous in
  time, sharing the same value  ${\sigma^2}^*_d$ of the variance.

  We consider the following hierarchical model
  \begin{eqnarray}
    \label{mod:var}
    &\;&y_t \mid \big(\mu,({\sigma^2}^*_{1}\dots {\sigma^2}^*_{|\rho|}),\rho \big)
    \stackrel{ind.}{\thicksim} N(\mu,\sigma_{t}^{2})~,\\
    &\;&\mu\mid \big(({\sigma^2}^*_{1}\dots {\sigma^2}^*_{|\rho|}),\rho \big)
    \thicksim N\left(m,\frac{\lambda_0}{T(\nu_0-1)}\right)~,\nonumber\\
    &\;& {\sigma^2}^*_{1}\dots {\sigma^2}^*_{|\rho|}\mid \rho
    \stackrel{i.i.d.}{\thicksim} IG(\nu_{0},\lambda_{0})~,\nonumber\\
    &\;&  \rho \sim \mbox{product distribution, with}\; \;
    C(S_d)=c \times (|S_d|-1)!~,\nonumber
    \end{eqnarray}
  with the variance of the Normal prior over $\mu$ equal to $\lambda_0/[T(\nu_0-1)]$,
  where $\lambda_0/(\nu_0-1)$ is the first moment of $IG(\nu_0,\lambda_0)$ and
  $1/T$  ia a scaling factor.

  The joint distribution is given by
  \begin{eqnarray*}
    &&f(\boldsymbol{y},\mu,\rho=
    \{S_{1},\dots,S_{|\rho|}\},\boldsymbol{\sigma^2})\propto\exp
    \left\{-\frac{T(\nu_0-1)}{2\lambda_0}(\mu-m)^2\right\}
    \\&\times&\exp\left\{-\frac{1}{2}\sum_{d=1}^{|\rho|}\sum_{t\in
      S_{d}}\frac{(y_t-\mu)^2}{\sigma_d^{2*}} -\sum_{d=1}^{|\rho|} \frac{\lambda_0}{\sigma^{2*}_d}\right\}
    \prod_{d=1}^{|\rho|} \left(\sigma^{2*}_d\right)^{-(\nu_0+1+|S_d|/2)}(|S_{d}|-1)!~.\nonumber
  \end{eqnarray*}

  In order to sample from the posterior distribution of the model
  and parameters we use a Gibbs algorithm that is a   generalization of the one used in the section
  \ref{s:means}. The algorithm consists of the three steps below.

  \begin{itemize}
  \item[\emph{Step} (i)~:] Sample $\mu$ from its full conditional distribution
    \begin{eqnarray*}
    \mu|\boldsymbol{\sigma^2},\boldsymbol{y}\sim N\left(\frac{m+\sum_{d=1}^{|\rho|}
    \frac{\lambda_0}{{T(\nu_0-1)\sigma^2}^{*}_d}\sum_{i\in S_d}y_i}{1+\sum_{d=1}^{|\rho|}
    |S_d|\frac{\lambda_0}{{T(\nu_0-1)\sigma^2}^{*}_d}},
    \frac{\frac{\lambda_0}{T(\nu_0-1)}}{1+\sum_{d=1}^{|\rho|} |S_d|\frac{\lambda_0}{{T(\nu_0-1)\sigma^2}^{*}_d}}\right)~.%\nonumber\\
    \end{eqnarray*}
  \item[\emph{Step} (ii)~:] Update each  $\sigma^2_{t}$,
    $t=1,\ldots,T$, by sampling  from the mixture
    \ben \label{eq:upsigma}
    &&\sigma^2_{t}|\boldsymbol{\sigma^2}_{-t},\boldsymbol{y}\sim\sum_{j\neq t}
    \tilde{q}_{tj}~\delta_{\sigma^2_{j}}(\sigma^2_{t})
    +\tilde{q}_{t0} \times IG\hspace{-0.1cm}\left(\nu_0+\frac{1}{2},\lambda_0+\frac{(y_t-\mu)^2}{2}\right),
    \een
    \noindent where $\boldsymbol{\sigma}^2_{-t}$ is obtained from
    $\boldsymbol{{\sigma}^2}$ by removing  $t$-th entry and
    $\delta_{\sigma^2_j}(\sigma^2_t)$ is the Dirac delta centered on $\sigma^2_t$.

    \noindent The distribution in equation~(\ref{eq:upsigma}) corresponds to a mixture of point masses
    and an Inverted Gamma distribution, with weights
    \begin{eqnarray*}
      &&\tilde{q}_{tj}\propto\frac{1}{\sqrt{\sigma_j^2}}\mathrm{e}^{-\frac{(y_{t}-\mu)^2}{2\sigma^2_{j}}},
      \\
      &&\tilde{q}_{t0}\propto
      c\times\frac{\Gamma\left(\nu_0+\frac{1}{2}\right)}{\Gamma(\nu_0)}
      \frac{2^{\nu_0+\frac{1}{2}}(\lambda_0)^{\nu_0}}{\left[(y_t-\mu)^2+2\lambda_0\right]^{\nu_0+\frac{1}{2}}}~,\\
      &&\sum_{j\neq t}
      q_{tj}+q_{t0}=1,
    \end{eqnarray*}
    where $\Gamma$ is the Euler Gamma function.
  \item[\emph{Step} (iii)~:] In order to  avoid being trapped in
    sticky regions of the Markov space, resample $\sigma^{2*}_{d}$
    from
    \be
    &&\sigma^{2*}_{d}\sim IG \hspace{-0.1cm}
    \left(\nu_0+\frac{\mid S_d\mid}{2},\lambda_0+\sum_{t\in
      S_d}\frac{(y_t-\mu)^2}{2}\right)\;\;\;d=1,\ldots,|\rho|.
    \ee
  \end{itemize}

  A well-known stylized fact about volatilities is the bursting
  effect and PPMs can be exploited to identify change
  points in volatility time series. This pro\-blem has been
  extensively considered by \cite{Loschi_etal:2003},
  \cite{Loschi_etal:2007} and \cite{Loschi_etal:2008}. Although we
  do not focus on this aspect here, in the empirical analysis in section
  \ref{s:numerical}  we shall use the results from the
  algorithm by \cite{Loschi_etal:2003},
  labelled $\boldsymbol{\sigma^2}$-CP,
  as a yardstick to be compared with our numerical results.

  \subsection{VaR Estimation \label{s:VAR}}

  We now present how the posterior distribution of VaR and consequently its Bayesian estimate can be obtained
  by using the output of the MCMC algorithms described in sections \ref{s:means} and \ref{s:variances}.

  First we focus our attention on the PPM  on the vector of means.
  Let indicate with
  $\boldsymbol{\mu}^*_{(\ell)}=\left({\mu_1^{*}}_{(\ell)},\dots,{\mu_{|\rho|}^{*}}_{(\ell)}\right)$
  and $\sigma^2_{(\ell)}$ respectively the  vector of  means
  and the  variance sampled at the $\ell$-th iteration of the Gibbs algorithm.
  At each iteration we obtain a peculiar clustering structure. All
  returns share the same value of $\sigma^2_{(\ell)}$, but each
  cluster is characterized by a different value
  ${\mu_d^*}_{(\ell)}$.
  In order to provide a single VaR estimate for each iteration  of the chain
  we propose to combine the different entries of
  $\boldsymbol{\mu}^*_{(\ell)}$ by means of an arithmetic average
  and we consider the following equation
  \ben\label{eq:VaRaverage}
  \frac{\Lambda_{(\ell)}}{W_0}=-\sum_{d=1}^{|\rho|}\frac{|{S_d}_{(\ell)}|}{T}{\mu_d^*}_{(\ell)}
  +\sigma_{(\ell)}\sqrt{2}\; \mathrm{erfc}^{-1}(2 \alpha)~.
  \een
  It is worth noticing that for trivial partitions, i.e. $|\rho|=1$,
  equation (\ref{eq:VaRaverage}) reduces to the usual expression
  given in equation~(\ref{eq:normalVaR}).

  If we impose  a clustering structure over the vector of variances,
  VaR can be computed in an analogous way   but the arithmetic
  average is performed over different values of
  ${\sigma^{*}_d}_{(\ell)}$,  that is
  \ben\label{eq:VaRaverage1}
  \frac{\Lambda_{(\ell)}}{W_0}=-{\mu_{(\ell)}}+\sum_{d=1}^{|\rho|}\frac{|{S_d}_{(\ell)}|}{T}\;{\sigma_d^{*}}_{(\ell)}\sqrt{2}\;
  \mathrm{erfc}^{-1}(2 \alpha)~.
  \een
  In this case all returns share the same value of $\mu_{(\ell)}$
  but each cluster is characterised by a different value of
  ${\sigma^{2*}_d}_{(\ell)}$.

  The resulting VaR estimate is obtained as the ergodic mean of the quantities $\Lambda_{(\ell)}$ in \eqref{eq:VaRaverage} or
  \eqref{eq:VaRaverage1} for $\boldsymbol{\mu}$-PPM or $\boldsymbol{\sigma^2}$-PPM respectively:
  \ben\label{eq:ergodicVaR}
  \frac{\Lambda}{W_0}\doteq \frac{1}{L}\sum_{\ell=1}^L  \frac{\Lambda_{(\ell)}}{W_0}.
  \een

  Finally, VaR under the $\boldsymbol{\sigma^2}$-CP model is computed in a similar way
  via equation (\ref{eq:ergodicVaR}).

  \section{Product Partition and Outliers Identification \label{s:Out}}

  PPMs can be a useful tool for outliers identification. Following
  \cite{Quintana_Iglesias:2003},  we work in a Bayesian decision
  theoretical framework and we propose an efficient algorithm for
  outliers identification.
  We model outliers as a shift in the mean of the data and consequently we fix our attention on
  $\boldsymbol{\mu}$-PPM. 
  The extent of that shift is indeed the criterion used by this model to induce a new cluster on the
  vector of returns, as emerge from the expression of the weights $q_{tj}$ and $q_{t0}$.

  Our aim is to select  the partition that best separates the main group of
  standard observations from one or more groups of atypical data.
  Each partition corresponds to a different model, and the best
  model is the one minimising a given loss function.
  Let $\left(\boldsymbol{\mu},\sigma^2\right)$ be the vector of
  parameters of the model and
  $\left(\boldsymbol{\mu}_{\rho},\sigma^2_{\rho}\right)$ the
  corresponding vector that results when fixing $\rho$. We consider
  the following loss function that combines the estimation of the
  parameters and the partition selection problems
  \begin{equation}\label{eq:loss}
    L(\rho,\boldsymbol{\mu}_{\rho},\sigma^2_{\rho},\boldsymbol{\mu},\sigma^2)
    =\frac{k_1}{T}\parallel\boldsymbol{\mu}_{\rho}-{\boldsymbol{\mu}}\parallel^2
    +k_2(\sigma^2_{\rho}-\sigma^2)^2+(1-k_1-k_2)|\rho|,
  \end{equation}
  where $\parallel\cdot\parallel$ is the Euclidean norm and $k_1$,
  $k_2$ are non-negative cost-complexity parameters with $k_1+k_2\leq
  1$.

  Minimising the expected value of (\ref{eq:loss}) is
  equivalent to choosing the partition  that minimises the following
  score function
  \begin{equation}\label{eq:score}
    SC(\rho)=\frac{k_1}{T}\parallel\hat{\boldsymbol{\mu}}_{B}(\boldsymbol{y})-
    \hat{\boldsymbol{\mu}}_{\rho}(\boldsymbol{y})\parallel^2
    +k_2\left[\hat{\sigma}_{B}^2(\boldsymbol{y})-\hat{\sigma}_{\rho}^2(\boldsymbol{y})\right]^2
    +(1-k_1-k_2)|\rho|,
  \end{equation}
  where the subscript ``$B$'' means that we consider the Bayesian
  estimates of the corresponding parameter whereas the subscript
  ``$\rho$'' indicates the estimate conditionally on a given
  partition $\rho$. Formally, we have that
  $\hat{\boldsymbol{\mu}}_B(\boldsymbol{y})=\uE{\boldsymbol{\mu}|\boldsymbol{y}}$,
  $\hat{\boldsymbol{\mu}}_\rho(\boldsymbol{y})=\uE{\boldsymbol{\mu}|\boldsymbol{y},\rho}$
  and analogously for $\hat{\sigma}_B^2(\boldsymbol{y})$ and
  $\hat{\sigma}_\rho^2(\boldsymbol{y})$. The Bayesian estimates are
  obtained via the Gibbs sampling algorithm described in the section
  \ref{s:means}. The evaluation of $\hat{\boldsymbol{\mu}}_\rho(\boldsymbol{y})$ and
  $\hat{\sigma}_\rho^2(\boldsymbol{y})$ also requires the use of the
  Gibbs sampling scheme, but in a structurally simpler version.
  Indeed the partition $\rho$ is fixed and we can sample from the
  joint posterior distribution performing iteratively \emph{Step}
  (i) and \emph{Step} (iii), but skipping \emph{Step} (ii).

  An exhaustive search
  on the space of all possible partitions is infeasible. In fact,
  for a set with $T$ elements, the number of all possible partitions is
  equal to $B(T)$, the \emph{Bell number} of order $T$,  recursively
  defined by $B(T+1)=\sum_{k=0}^T \binom{T}{k}B(k)$, with $B(0)=1$.
  This quantity is extremely large even for moderate values of $a$,
  therefore we  need to restrict our search to a tractable subset of
  all partitions.

  In order to find the minimum of the score function in
  equation~(\ref{eq:score}), we perform an exhaustive search over the
  partitions with cardinality up to three, selected as follow.

  \begin{itemize}
  \item[i)] Let $\boldsymbol{\mu}_B=(\mu_1,\ldots,\mu_T)$ be the
    vector of the Bayesian estimates of the returns means,
    and $\boldsymbol{\widetilde{\mu}}_B=({\widetilde{\mu}}_1,\ldots,{\widetilde{\mu}}_{\widetilde{T}})$
    be the vector  of the unique entries of
    $\boldsymbol{\mu}_B$ sorted in increasing order, with
    $\widetilde{\mu}_1\doteq\mathrm{min}~(\mu_t)$,
    $\widetilde{\mu}_{\widetilde{T}}\doteq\mathrm{max}~(\mu_t)$,
    and  $\widetilde{T}\leq T$.
  \item[ii)] We perform our search of
    the optimal partition over the set of the partitions
    $\rho=\left\{S_1,S_2,S_3\right\}$,
    where
    $S_1=\left\{t:\mu_t < \widetilde{\mu}_i\right\}$,
    $S_2=\left\{ t:\widetilde{\mu}_i \leq \mu_t \leq
    \widetilde{\mu}_j \right\}$,
    and
    $S_3=\left\{t:\mu_t > \widetilde{\mu}_j\right\}$ with $i,j=1,\ldots,\widetilde{T}$.
    We select as optimal partition the one for which the score function
    achieves the minimum value in~\eqref{eq:score}.
  \end{itemize}

  When $i\neq 1$ and $j\neq \widetilde{T}$, $\rho$ is a genuine
  cardinality-3 partition. The indexes in $S_1$ and $S_3$ may be considered as
  representative of those returns being in the ``left tail'' and the ``right tail''
  of the empirical distribution of $\boldsymbol{y}$.
  $S_2$ corresponds to elements occupying the central region of this distribution.
  When $i=1$ and $j=\widetilde{T}$, we are exploring the trivial partition, $S_1=S_3=\emptyset$.
  If $i=1$ or $j=\widetilde{T}$, partitions have just two clusters. However, there is an alternative way
  to generate cardinality-2 partitions. Given every cardinality-3 partition $\rho$, we consider the new
  partition $\hat{\rho}=\left\{S_1,S_2 \right\}$, with
  $S_1\doteq S_1 \cup S_3$.
  This step is necessary for our search to be exhaustive also over the space of cardinality-2 partitions.

  Once the optimal partition has been found, we identify the outliers with those
  elements in $\boldsymbol{y}$ whose indexes belong to the sets with lowest cardinality.

  \section{Empirical Analysis of Financial Data}\label{s:numerical}

  \subsection{The Data}

  The methodologies described in the previous sections are now
  illustrated and tested over the MIB30 index and its three
  components with the highest excess of kurtosis,
   where standard approaches based on Normal distributions usually fail.
  In particular we apply our analysis to the Italian assets Lottomatica (LTO.MI),
  Mediobanca (MB.MI) and Snam Rete Gas (SRG.MI). We consider time
  series  of daily returns from April 2004 to March 2008.
  All time series are made of
  1000 daily returns. The data are freely downloadable from the site
  \emph{http://it.finance.yahoo.com}.

  \subsection{Choice of Hyperparameters  and Computational Details}

  In the examples below we use the following values of the
  hyperparameters.
  In models (\ref{mod:mean}), (\ref{mod:mean-ger}) and (\ref{mod:var})
  we set $m=0$, while $\tau_0^2=10^3$ in (\ref{mod:mean}). The choice
  for $m$ can be motivated by the fact that  in VaR estimation for short time
  horizon, typically from one day until one week,
  the value of the mean is usually neglected, see e.g.
  \cite{Mina_Xiao:2001}.
  In the Inverted Gamma distribution we set
  $\lambda_0=0.0101$ and $\nu_0=2.01$. With these choices we have
  prior expectation and variance $0.01$ for $\sigma^2$,
    reflecting what is known from the past experience about the
  volatility behaviour for equity assets.
  The value of $c$ that
  controls the clustering structure over the vector of parameters is
  set to 1, in order to favour the creation of a small number of large
  clusters. As far as concern the score function parameters of
  equation (\ref{eq:score}) we set $k_1 \sim 0.996 $ and $k_2 \sim  0.002$,
  giving priority to the estimation of $\boldsymbol{\mu}$,
  and imposing little restriction on the estimation of the other parameters.
  For the
  $\boldsymbol{\sigma^2}$-CP model that we use as yardstick  model we
  set the priors' parameters following the suggestions given in
  \cite{Loschi_etal:2003}. In particular we consider the conjugate
  Normal-Inverted-Gamma model, with the probability $p$ that a change
  occurs at any instant in the sequence equal to 0.1.

  The programs are written in Fortran 77 language,
  with basic function of linear algebra provided by BLAS and SLATEC
  libraries. Random number generators, Normal and Gamma sampling are
  based on the algorithms implemented in the RANDOM library. The
  interested reader can download them and find more detail browsing
  the Netlib repository at \emph{http://www.netlib.org}.
  The algorithm proposed by \cite{Loschi_etal:2003} is freely available
  at the web site \linebreak \emph{ftp://ftp.est.ufmg.br/pub/loschi/}.

  We run the MCMC algorithms with 10000 sweeps and a burn-in equal to
  1000. Convergence of the MCMC algorithm
  is assessed using diagnostics implemented in the
  package BOA, see \cite{Smith:2001}.  All the numerical computations
  are performed with an AMD Athlon 64 X2 3800 2.0 GHz processor and
  2.0 GByte of RAM, OS Gentoo Linux kernel 2.6.22. Each program
  takes nearly 15 minutes to generate the ergodic sample and to
  compute the parameters posterior distributions. The clustering
  structure for each step of the chain and the relative frequencies
  of the partitions are computed by means of sorting algorithms. It
  takes further 10 minutes to accomplish this task. In our programs
  we use sorting algorithms implementing strategies of $O(T^2)$
  computational complexity. It is possible to reduce the
  computational burden by means of $O(T\log{T})$ algorithms. However
  it is crucial that sorting preserves relative order of records
  with equal keys, but this in general requires storage of an
  auxiliary amount of memory.

  \subsection{VaR Results}

  In table \ref{tab:VaR} we report Bayesian estimates  of percentage VaR for
  $\alpha=1\%$ and $\alpha=5\%$ and the $68\%$ posterior credible
  interval.

  \begin{center}
    TABLE \ref{tab:VaR} ABOUT HERE
  \end{center}

  The estimates of  VaR obtained with
  $\boldsymbol{\sigma^2}$-PPM and $\boldsymbol{\sigma^2}$-CP are in
  good agreement even if the two approaches are quite different in
  spirit.  The former approach is a natural extension of the
  $\boldsymbol{\mu}$-PPM to the vector of variances while the latter
  one is specific for  change point
  identification.

  The PPM on the vector of means in general underestimates VaR
  with respect to the values given by the PPMs applied to the
  variances. This fact can be empirically justified noticing that
  for daily time horizons the contribution to VaR due to the
  volatility $\sigma$ is of order ten greater than that due to the
  mean $\mu$.

\begin{figure}[p!]
    \caption{\label{fig:posteriors} VaR posterior distribution for $\alpha=1\%$.}
    \vspace{0.5cm}
    \centering {\large{$\boldsymbol{\mu}$-PPM}}\\\vspace{0.5cm}
    \includegraphics[width=0.235\textwidth]{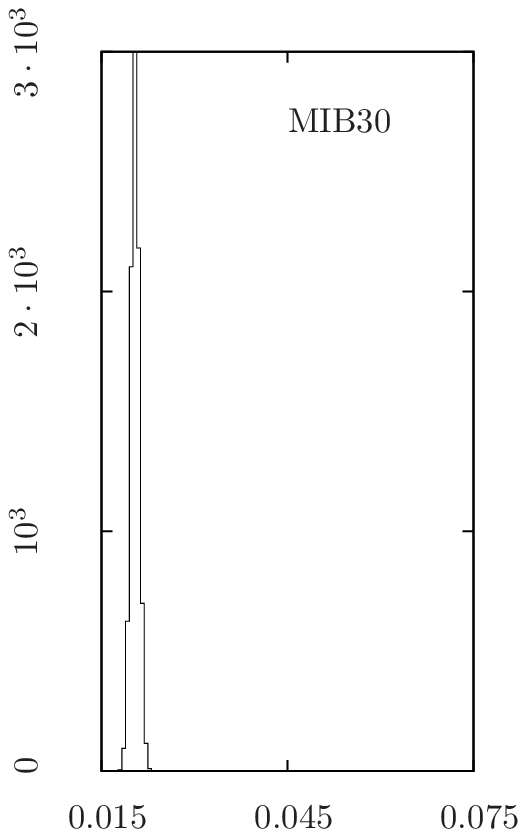}
    \hfill
    \includegraphics[width=0.235\textwidth]{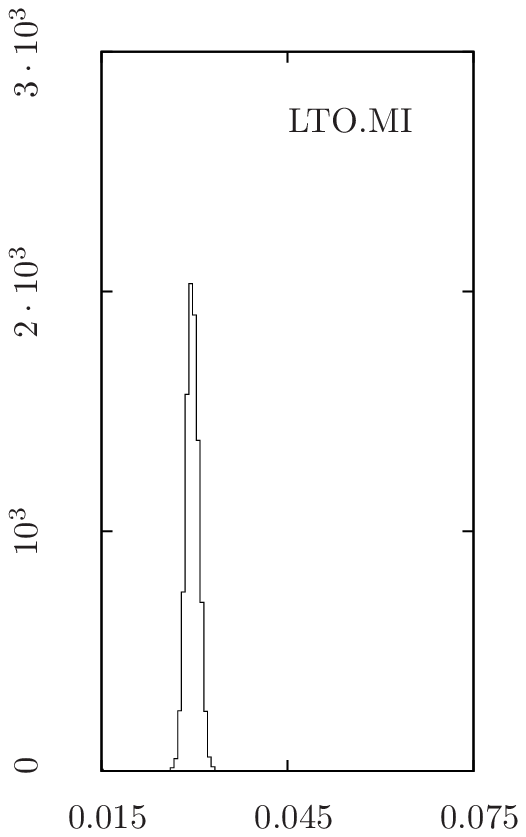}
    \hfill
    \includegraphics[width=0.235\textwidth]{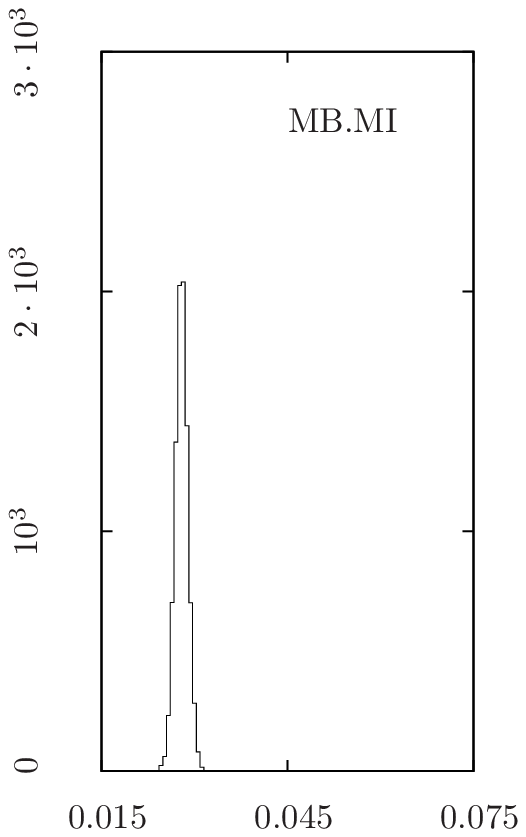}
    \hfill
    \includegraphics[width=0.235\textwidth]{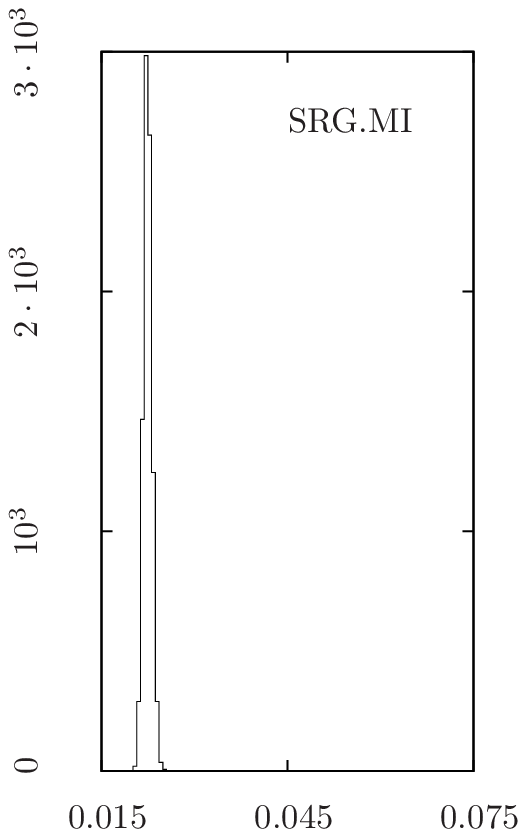}\\ \vspace{0.5cm}
    \centering {\large{$\boldsymbol{\sigma^2}$-PPM}}\\ \vspace{0.5cm}
    \includegraphics[width=0.235\textwidth]{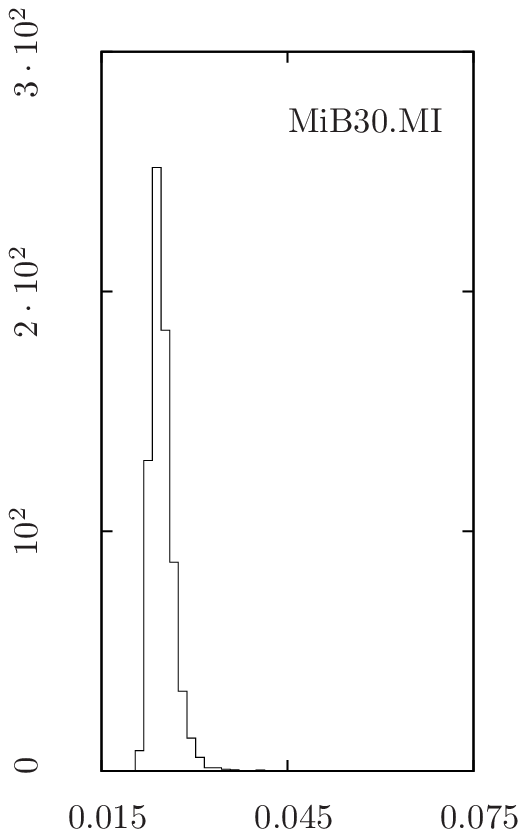}
    \hfill
    \includegraphics[width=0.235\textwidth]{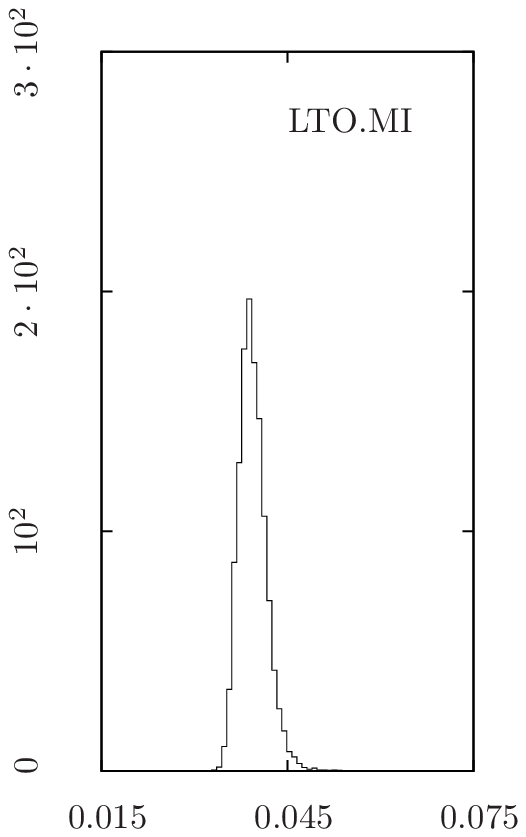}
    \hfill
    \includegraphics[width=0.235\textwidth]{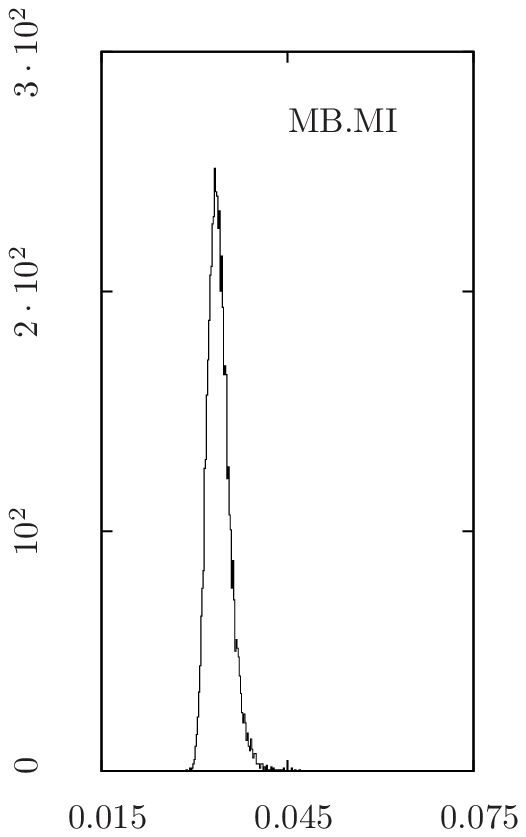}
    \hfill
    \includegraphics[width=0.235\textwidth]{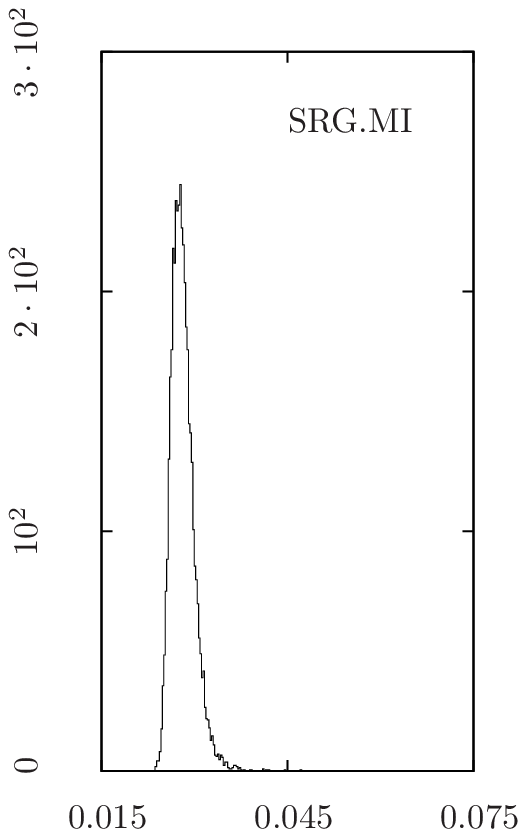}
  \end{figure}

  Figure~\ref{fig:posteriors} depicts   posteriors distributions for
  VaR estimates at level $\alpha=1\%$. In the first row we present
  the results based on the $\boldsymbol{\mu}$-PPM approach, while
  the second corresponds to $\boldsymbol{\sigma^2}$-PPM. The
  posterior distribution of VaR presents a  higher variability under
  the $\boldsymbol{\sigma^2}$-PPM  approach than under the
  $\boldsymbol{\mu}$-PPM VaR one.

  The posterior expectation
  of the number of clusters is low for both the $\boldsymbol{\mu}$-PPM and
  $\boldsymbol{\sigma^2}$-PPM approaches and, moreover, the partitions are
  characterised by a very large cluster and few small ones.
  The results are presented in table \ref{tab:clusters}.

  \begin{center}
    TABLE \ref{tab:clusters} ABOUT HERE
  \end{center}

  The arithmetic average in equations~(\ref{eq:VaRaverage}) and (\ref{eq:VaRaverage1})
  is therefore dominated by the values of ${\mu^*_d}_{(\ell)}$ and
  ${\sigma^{2*}_d}_{(\ell)}$ that correspond to the largest
  cluster, while outlying clusters introduce corrections to VaR.

  \begin{center}
    FIGURE \ref{fig:posteriors} ABOUT HERE
  \end{center}

\begin{figure}[p] \caption{\label{fig:MLvsBayesian}
      Comparison between classical and Bayesian estimates of VaR(\%).
      We consider two values for the  levels $\alpha=1\%$ (top panel)
      and $\alpha=5\%$ (lower panel).
    }
    \vspace{0.5cm}
    \begin{minipage}[b]{1.\textwidth}
      \begin{center}
    \includegraphics[scale=1.2]{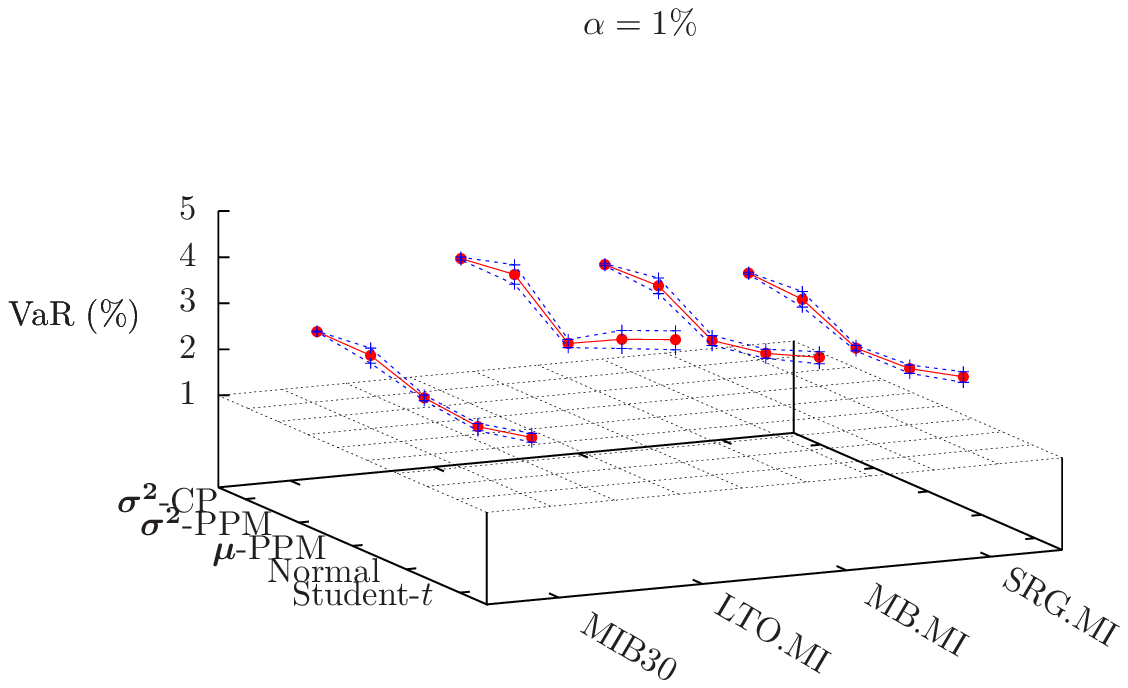}
      \end{center}
    \end{minipage}
    \begin{minipage}[b]{1.\textwidth}\vspace{0.5cm}
      \begin{center}
    \includegraphics[scale=1.2]{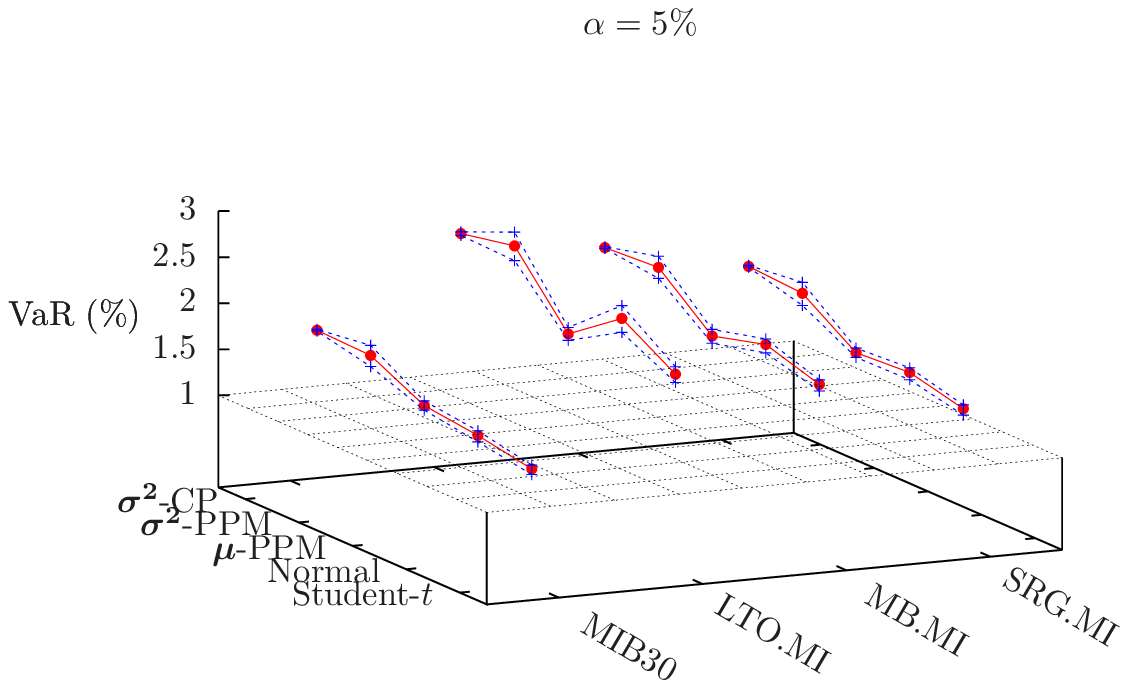}
      \end{center}
    \end{minipage}
  \end{figure}

  We now compare our results with those obtained  with standard
  parametric approaches based on ML estimators for the mean and variance. 
  In particular we consider results obtained with a Normal model and with the generalised
  Student-$t$ (GST) distribution, see e.g. \cite{Bormetti:2007}. In the
  GST we set the tail index $\nu>2$, in order to keep
  the variance finite, see last column of table \ref{tab:MLVaR}.
  In the following we consider the GST as the benchmark for our analysis since it presents a
  good agreement with historical simulations, see
  \cite{Bormetti:2007}. 
  For the daily returns under study we report in figure~\ref{fig:MLvsBayesian} 
  ML estimates and their $68\%$ confidence intervals computed from the cumulative function obtained
  generating 1000 bootstrap copies of the original time series. 
  Numerical details are reported in tables \ref{tab:VaR} and \ref{tab:MLVaR}.
  The solid line in figure~\ref{fig:MLvsBayesian} joints the estimated values of VaR(\%) while
  the dashed lines connects  the boundaries of the 68\%
  credible/bootstrap intervals.

  \begin{center}
    TABLE \ref{tab:MLVaR} ABOUT HERE
  \end{center}

  \begin{center}
    FIGURE \ref{fig:MLvsBayesian} ABOUT HERE
  \end{center}

  At $\alpha=1\%$  the  results obtained with
  $\boldsymbol{\sigma^2}$-PPM and $\boldsymbol{\sigma^2}$-CP are the
  ones in best agreement with the GST
  distribution, while Normal and $\boldsymbol{\mu}$-PPM
  underestimate VaR. The situation is different if we consider
  $\alpha=5\%$. In this case $\boldsymbol{\mu}$-PPM is the only one
  in agreement with the GST distribution, while
  $\boldsymbol{\sigma^2}$-PPM and $\boldsymbol{\sigma^2}$-CP
  overestimate VaR.

  For time horizons longer than one-day, we focus mainly on the 10-day holding period,
  as required by Basel Committee for the computation of regulatory VaR. 
  Indeed, the Committee prescribes the following formula for the calculation
  of regulatory capital for market risk
  \begin{equation*}%\label{eq:LRC}
   MRC_t=\max\left(\frac{h}{60}\sum_{i=1}^{60}\Lambda^{0.01}_{t-i}(10),\Lambda^{0.01}_{t}(10)\right),
  \end{equation*}
  where $MRC_t$ is the market risk  capital at time $t$, $\Lambda^{0.01}_{t-i}(10)$ is VaR 
  (not normalized by $W_0$, see equation (\ref{eq:normalVaR})) at $\alpha=1\%$ for 10-day ahead computed using past returns up to time
  $t-i$ and $h$ is a penalty multiplier ranging from 3 to 4, fixed according to the traffic light rule, see \cite{Basel:2006}
  for more details.
  From the original daily returns time series we compute the series of non-overlapping 10-day returns
  to have the 10-day VaR forecast for both $\boldsymbol{\mu}$-PPM and $\boldsymbol{\sigma^2}$-PPM.
  This approach can be easily generalised to an arbitrary holding period.
  \begin{center}
    TABLE \ref{tab:10MLVaR} ABOUT HERE
  \end{center}
  In table \ref{tab:10MLVaR} we report the estimated 10-day ahead VaR(\%) for
  the standard significance levels $\alpha=1\%$ and $\alpha=5\%$ with their $68\%$
  credible intervals. Basel regulations recommend to use the so-called square-root-of-time
  rule to obtain the 10-day VaR from the one-day VaR. However, as already pointed out in
  \cite{Danielsson:1998}, this rule strongly depends on the assumption of normally i.i.d. returns.
  The ratio between 10-day and one-day VaR estimated with parametric PPMs is readily computed and
  indeed our results confirm a statistically significative violation of the square root scaling law,
  thus highlighting the ability of our approach to better capture the properties of returns time series.

  \subsection{Sensitivity Analysis and Outliers Detection}

  Although the choices of parameters value used in the previous
  sections represent our prior knowledge and beliefs about the
  problem, it is illustrative to assess the sensitivity of the
  results to other choices of the hyperparameters.

\begin{figure}[p]\caption{
     \label{fig:CvsVaR} Sensitivity of $\alpha=1\%$ VaR(\%) estimates for the $\boldsymbol{\mu}$-PPM model with respect to the value of the
      hyperparameter $c$ in the cohesion function (\ref{eq:cohesion}).
      The other hyperparameters assume the values quoted in the
      main text.
    }
    \vspace{0.5cm}
    \centering
    \includegraphics[width=.95\textwidth]{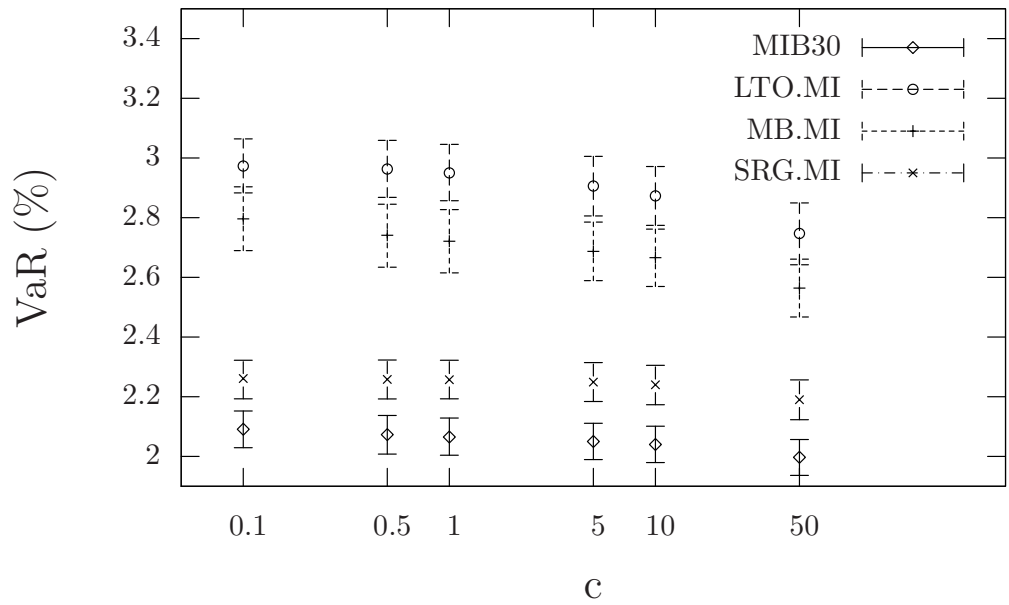}
  \end{figure}

  We first consider the dependence of VaR estimates on the value of
  the $c$ in the cohesion function in (\ref{eq:cohesion}).
  In  figure~\ref{fig:CvsVaR} we plot the results for the $\boldsymbol{\mu}$-PPM model
  for $\alpha=1\%$ and $c=0.1,0.5,1,5,10,50$.

  \begin{center}
    FIGURE \ref{fig:CvsVaR} ABOUT HERE
  \end{center}

  Note that for MIB30 and SRG.MI, the results are remarkably robust
  for  a wide range of values of $c$. For MB.MI and LTO.MI the estimated value
  of VaR exhibits a slightly decreasing trend. 

\begin{figure}[p]\caption{
      \label{fig:AvsVaR} Sensitivity of $\alpha=1\%$ VaR(\%) estimates with respect to the value of the
      hyperparameters $\lambda_0=a(a+1)$ and $\nu_0=2+a$. The other hyperparameters assume the values quoted in the
      main text.
    }\vspace{0.5cm}
    \centering{\large{$\boldsymbol{\mu}$-PPM} }\vspace{0.3cm}
    \includegraphics[width=.95\textwidth]{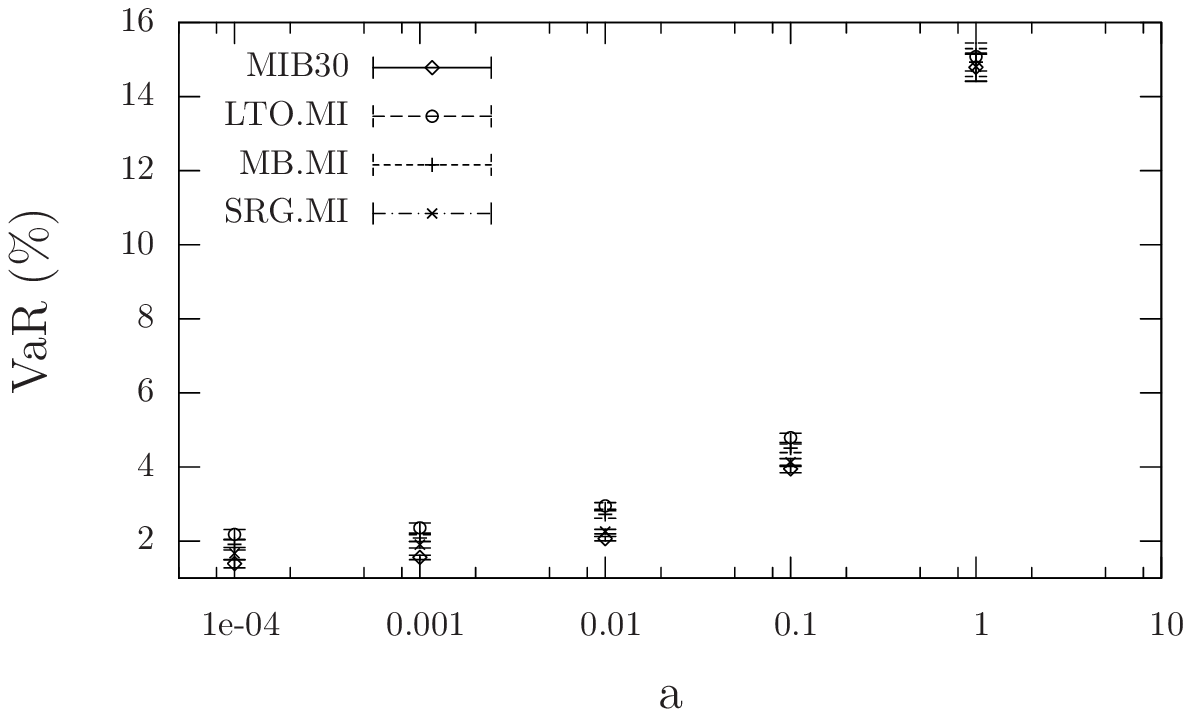}\\
    \centerline{\large{$\boldsymbol{\sigma^2}$-PPM}}\vspace{0.3cm}
    \includegraphics[width=.95\textwidth]{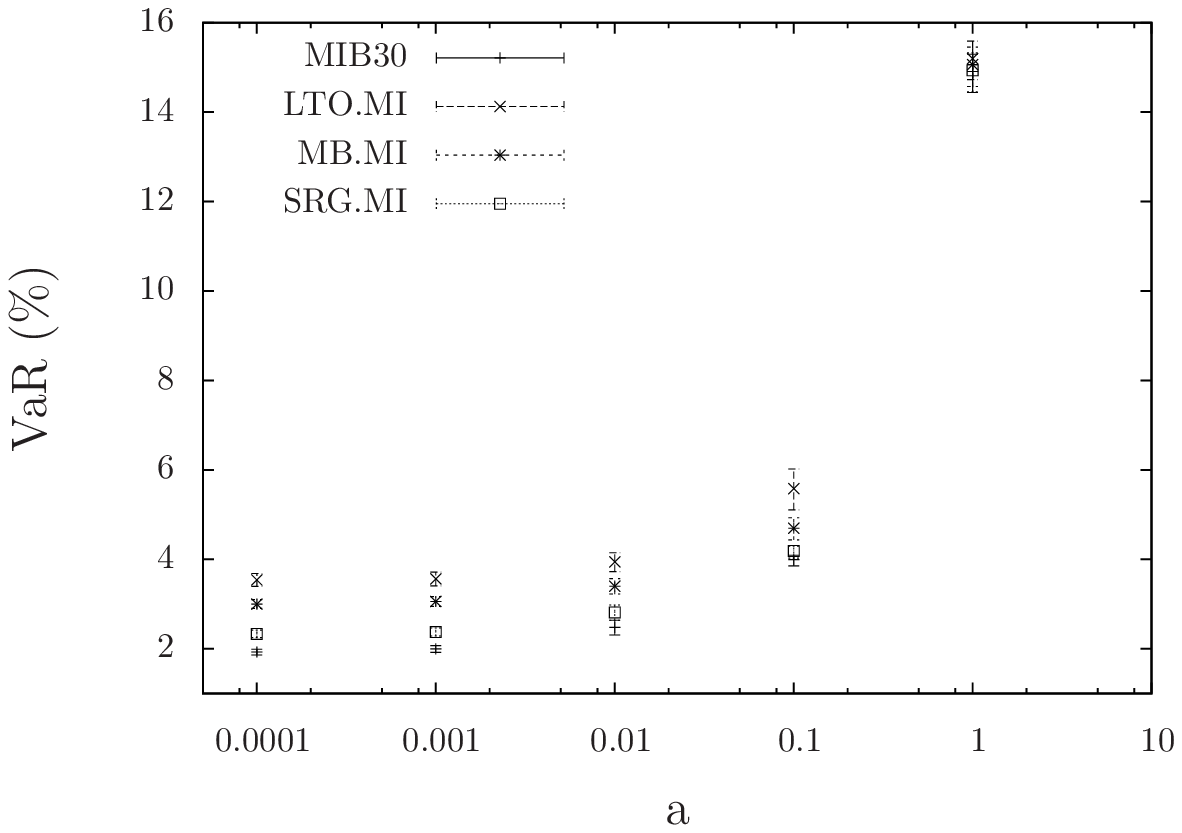}
  \end{figure}

  To study the sensitivity of our results to the parameters
  $\lambda_0$ and $\nu_0$ of the Inverted Gamma distribution it is
  convenient to re-express them in terms of a common parameter  $a$.
  We set $\lambda_0=a(a+1)$ and $\nu_0=2+a$ in order to obtain prior
  expectation and variance for $\sigma^2$ both equal to $a$.
  In figure~\ref{fig:AvsVaR} we present the results for
  $a=0.0001,0.001,0.01,0.1,1$.
  For  $a=1$ we have completely out-of-scale results.

  \begin{center}
    FIGURE \ref{fig:AvsVaR} ABOUT HERE
  \end{center}

\begin{figure}[p]\caption{
      \label{fig:LTOAvsVaRdistr} Posterior distributions for VaR at
      level $\alpha=1\%$ for Lottomatica as a function of $a$.
      The other hyperparameters assume the values
      quoted in the main text.
    }
    \vspace{0.5cm}
    \centering{\large{$\boldsymbol{\mu}$-PPM}}\\\vspace{0.3cm}
    \includegraphics[width=.85\textwidth]{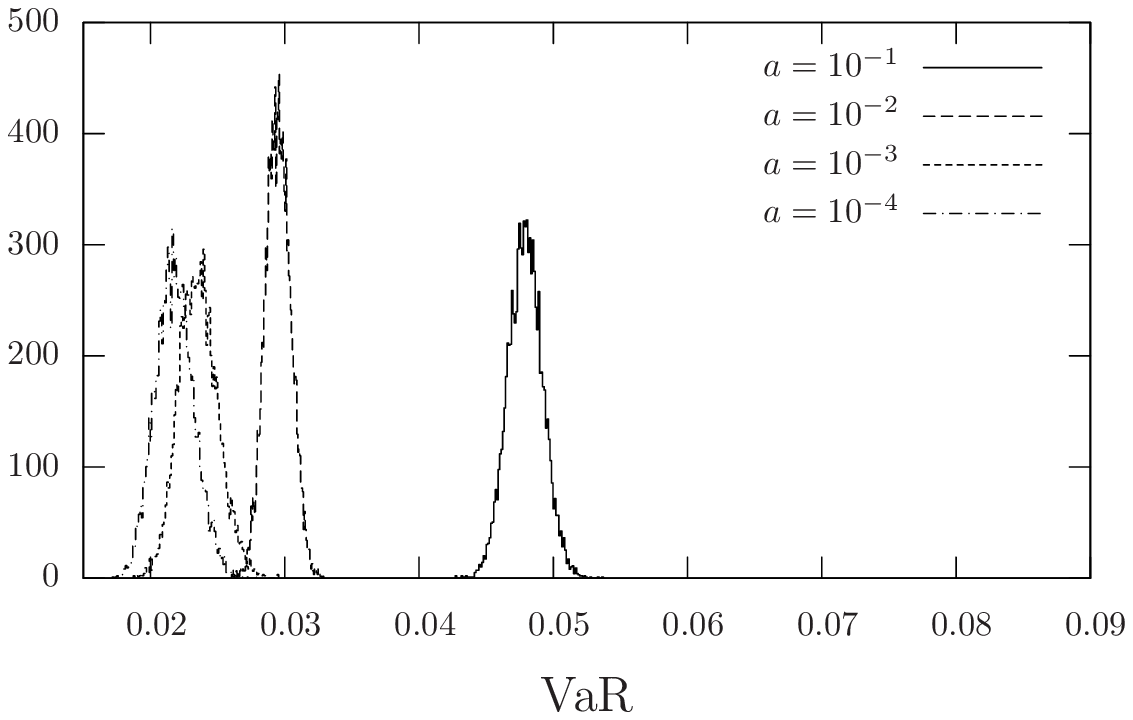}\\
    \vspace{0.3cm}
    \centerline{\large{$\boldsymbol{\sigma^2}$-PPM}}\vspace{0.3cm}
    \includegraphics[width=.85\textwidth]{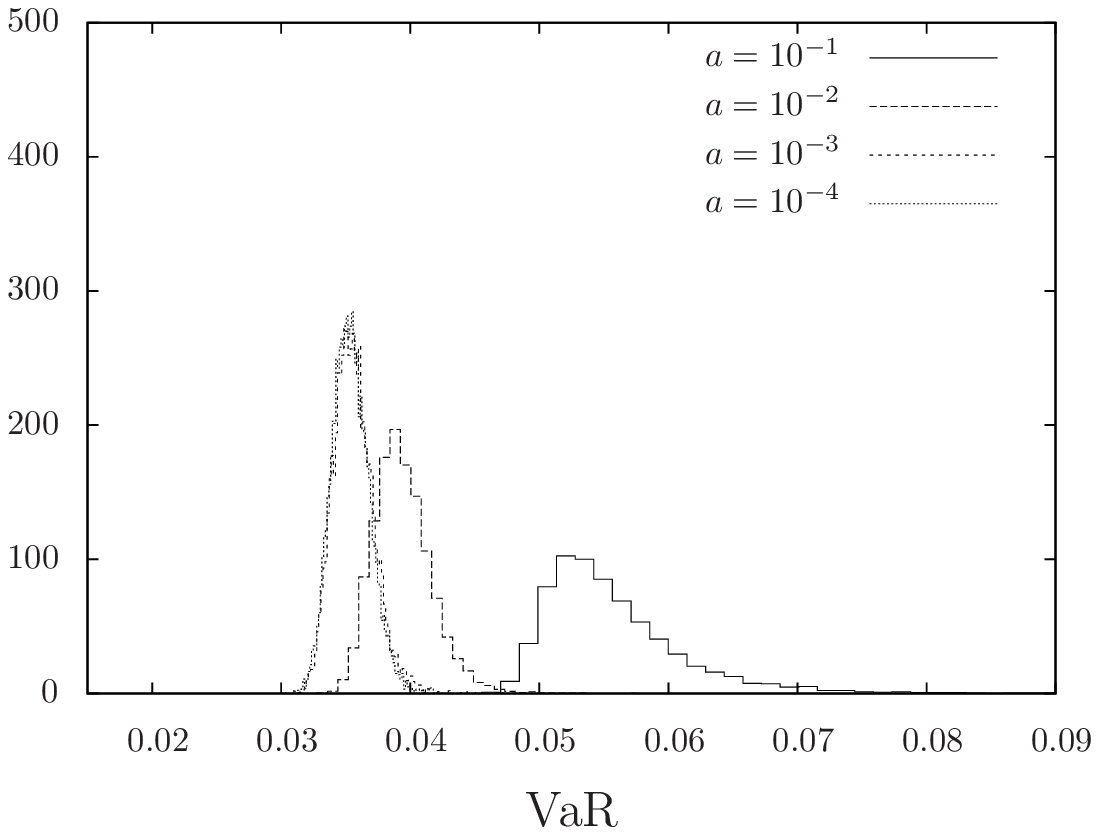}
  \end{figure}

  In this paper we use $a=0.01$, reflecting past knowledge
  regarding the problem at hand.
  For this reason we focus on a region around this value.
  The highest stability is reached when the PPM approach is applied
  to the vector of variances. In fact for $a\leq 0.01$ the results within the
  68\% credible intervals are almost identical. The
  $\boldsymbol{\mu}$-PPM is less stable. These results are confirmed
  in figure~\ref{fig:LTOAvsVaRdistr} where we plot the posterior
  distributions for LTO.MI   $\alpha=1\%$ VaR, with
  $a=0.0001,0.001,0.01,0.1$.

  \begin{center}
    FIGURE \ref{fig:LTOAvsVaRdistr} ABOUT HERE
  \end{center}

  We note that for $a=0.0001$ and $a=0.001$ the distributions
  obtained with the $\boldsymbol{\sigma^2}$-PPM are almost
  overlapping. A similar behaviour is observed for the other three
  time series.

  We also explored separately the role played by $\lambda_0$ and
  $\nu_0$ and we found that $\lambda_0$ assumes a crucial role. We
  than tested the effects of an hyperprior over the scale
  parameter $\lambda_0$. We considered various combination of
  the $\eta$ and $\phi$ parameters, as given in
  equation~(\ref{eq:hyperlambdafullcond}).  For all the tested values
  we were not able to achieve a reasonable sensitivity
  reduction. For the sake of parsimony
  we do not report here our results.

\begin{figure}[p]\caption{\label{fig:outliers} Detected outliers.}
    \vspace{0.2cm}
    \centering
    \includegraphics[width=0.49\textwidth]{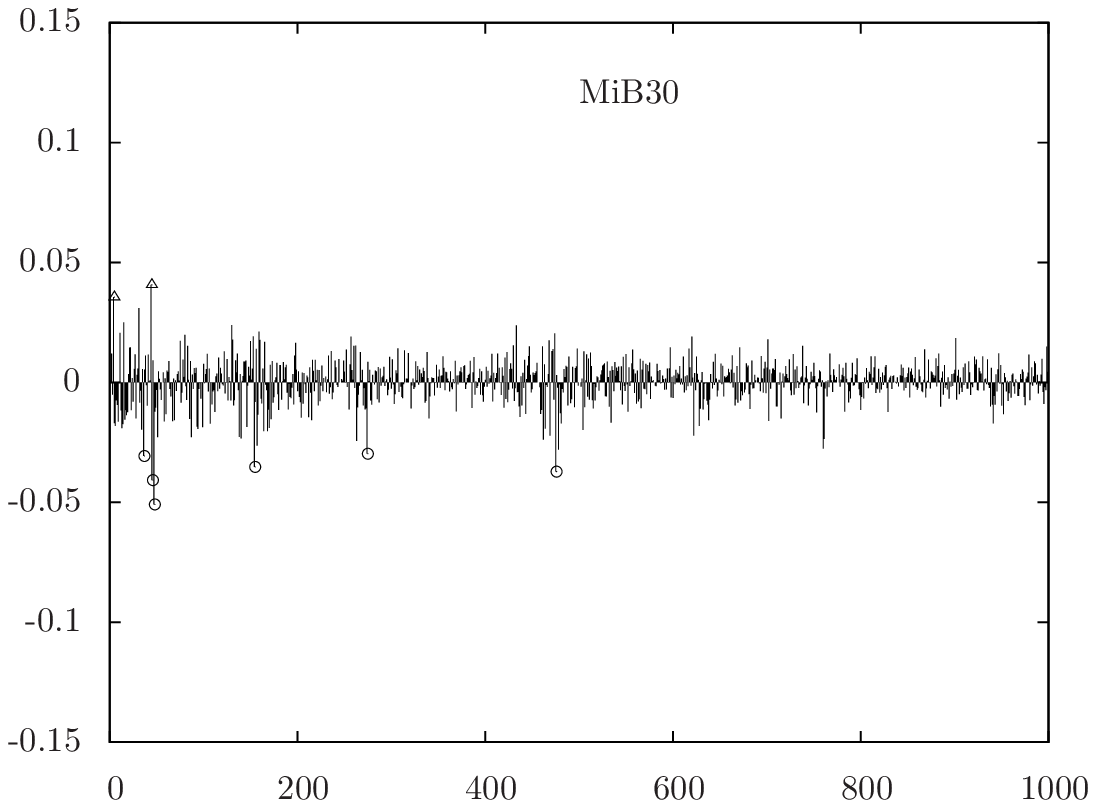}
    \hfill
    \includegraphics[width=0.49\textwidth]{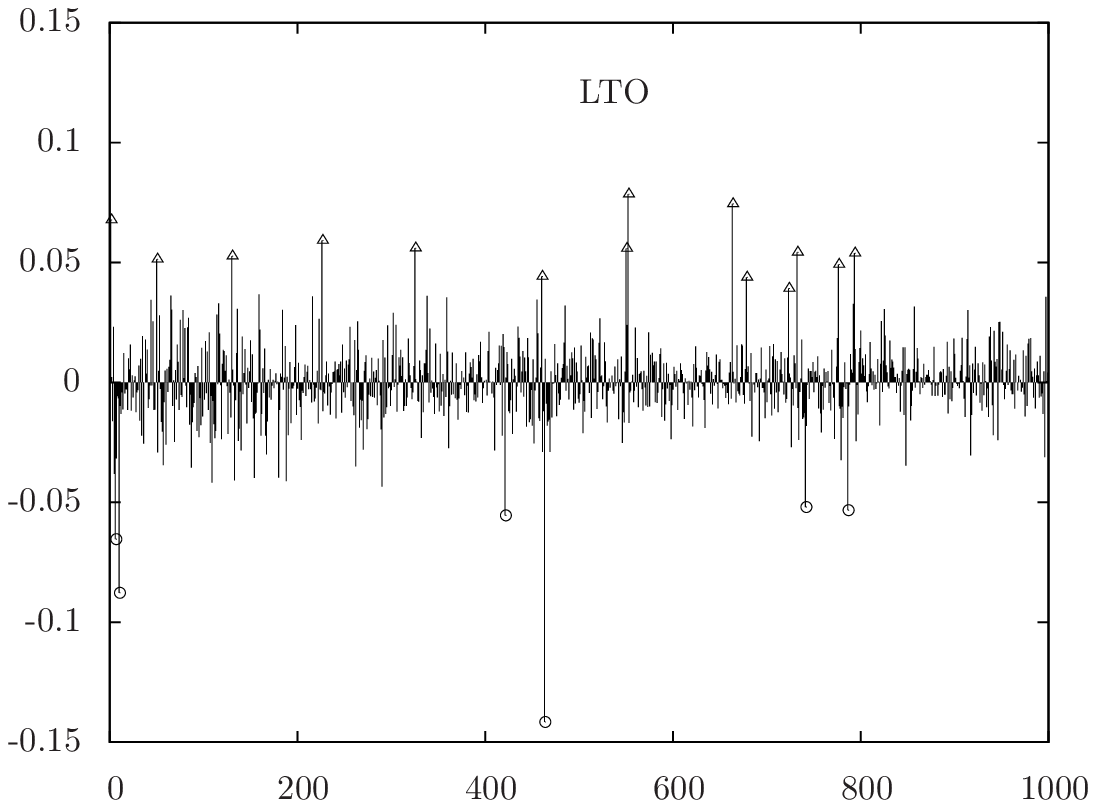}\\
    \includegraphics[width=0.49\textwidth]{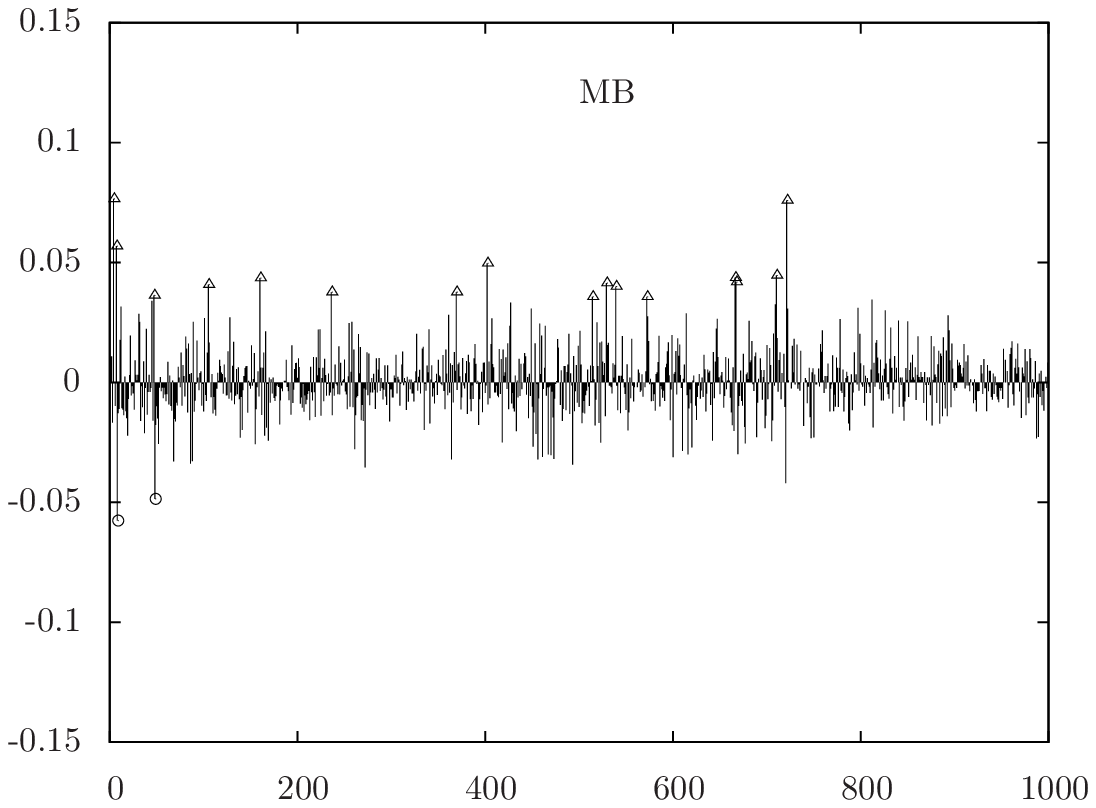}
    \hfill
    \includegraphics[width=.49\textwidth]{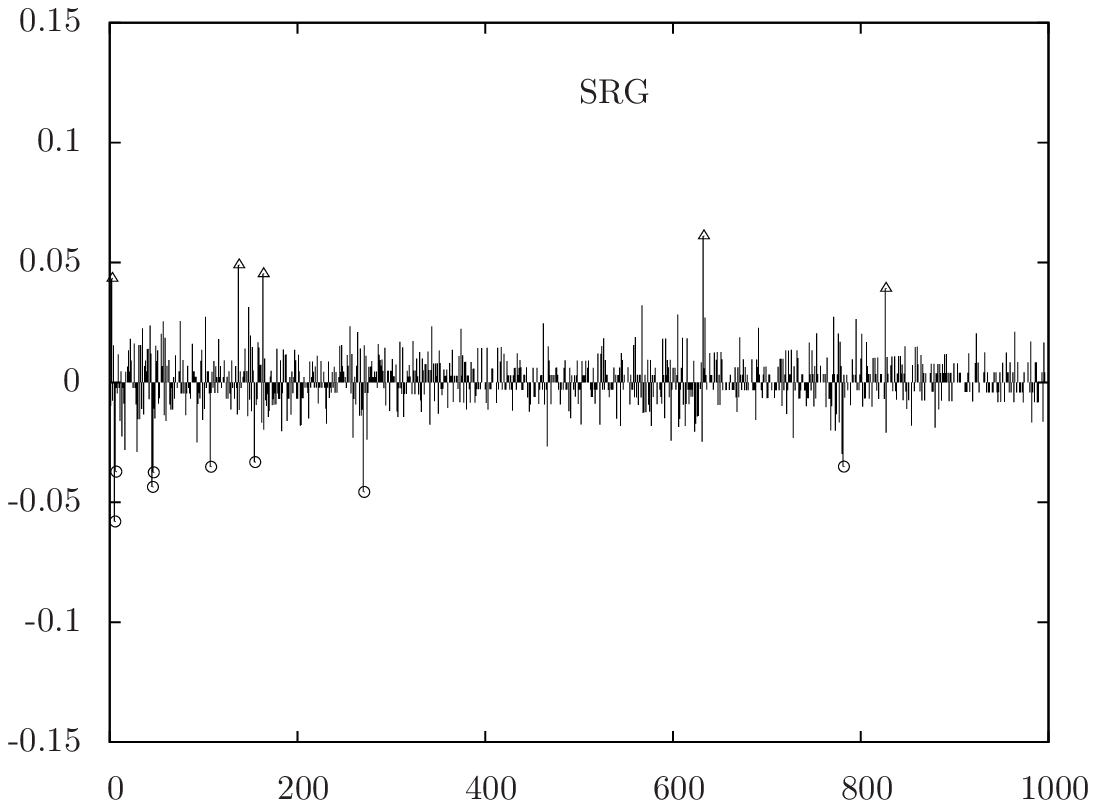}
  \end{figure}

  \begin{center}
    FIGURE \ref{fig:outliers} ABOUT HERE
  \end{center}

  In order to identify outlying points we apply  the procedure
  described in section \ref{s:Out}. The results are reported in
  figure~\ref{fig:outliers}.
  Returns corresponding to atypical values are represented by a
  small triangle (gains) or a small circle (losses).
  Their identification represents a by-product result of our
  approach to VaR computation. It could be interesting to investigate
  the economical reasons responsible for the anomalous fluctuations of assets price,
  along the same lines depicted in \cite{DeGiuli_etal:2009}.
  Finally it is interesting to investigate the stability of our procedure with
  respect to the value of $c$. Table~\ref{tab:outliers} summarizes the results of our analysis
  when increasing $c$ from 0.1 until 50.

  \begin{center}
    TABLE \ref{tab:outliers} ABOUT HERE
  \end{center}

  The outliers identification algorithm appears to be quite stable.
  As expected, on average the number of outlying points increases for increasing values of $c$.

  \subsection{Backtesting Procedures}

  The current internal model verification procedure of the Basel II
  framework consists of recording the daily exceptions of
  the $1\%$ VaR over the last year.
  We apply standard coverage tests to assess the accuracy of
  our VaR model; in particular we consider the unconditional
  coverage (UC)
  test by \cite{Kupiec:1995} and the conditional coverage (CC) one by
  \cite{Christoffersen:1998}. Kupiec's test focuses on whether the
  actual number of VaR exceptions is equal to their expected
  number. Assuming that the probability of observing an
  exception is $p$, the number of exceptions out of a sample of $N$
  observation follows a Binomial distribution $Bin(N,p)$.
  The null hypothesis $p=\alpha$ can be assessed by using the following generalised
  likelihood ratio test
  \begin{equation*}
     LR_{UC}  = -2\log\left[ (1-\alpha)^{N-n}\alpha^{n} \right]
     + 2 \log \left[ (1-n/N)^{N-n}(n/N)^{n} \right]
  \end{equation*}
  where $n$ is the observed number of exceptions. This quantity is asymptotically
  distributed chi-square with one degree of freedom under the null hypothesis,
  and allows us to reject the model
  at $5\%$ significance level when $LR_{UC} > 3.84$.
  The $LR_{UC}$ can be extended to test the serial independence of deviations,
  introducing a deviation indicator which is equal to 0 if VaR is not exceeded and
  1 otherwise. We consider the following combined test statistics (Christoffersen's test)
  \begin{align*}
     LR_{CC}&=LR_{UC}+ LR_{IND}\\
     LR_{IND}&=-2 \log \left[ (1-n/N)^{N_{00}+N_{10}}(n/N)^{N_{01}+N_{11}} \right]\\
     &+2 \log \left[ (1-\pi_0)^{N_{00}}\pi_0^{N_{01}}(1-\pi_1)^{N_{10}}\pi_1^{N_{11}} \right]
  \end{align*}
  where $N_{ij}$ is the number of days in which state $j$ occurred in one day while was $i$
  the previous day, and $\pi_i$ the probability of observing an exception conditional
  on the state $i$ the previous day, that is $\pi_0=N_{01}/(N_{00}+N_{01})$ and
  $\pi_1=N_{11}/(N_{10}+N_{11})$. The null hypothesis for the independence test
  states that the violation occurred one day does not depend upon the indicator state the previous day.
  Under this hypothesis, the $LR_{CC}$ statistics is distributed chi-square with two degrees of
  freedom and the VaR model will be rejected at $5\%$ significance level if $LR_{CC}>5.99$.

  We perform the validation tests
  described above to all our data series, using a rolling window of returns
  to compute the VaR estimate by our models and comparing this estimate
  with the realized return. More precisely, at each stage $J=1,\ldots,N$,
  our Gibbs sampling algorithms compute the ex ante VaR estimate $VaR^{\alpha}_{MAX_J}$ 
  using the returns $y_{i}$ with $i=J,\ldots,MAX_J$;
  then we check $VaR^{\alpha}_{MAX_J}$ against the ex post realized return $y_{MAX_J+1}$.
  An exception occurs when $y_{MAX_J+1} < - VaR_{MAX_J}^{\alpha}$. A state indicator $I_J$ is
  set equal to 1 if we register an exception, and equal to 0 otherwise.
  This way we obtain the numbers $n$ and $N^{ij}$ needed to compute the $LR_{CC}$ statistics.

  \begin{center}
    TABLE \ref{tab:backtest} ABOUT HERE
  \end{center}

  Choosing $MAX_J=J+744$, we are able to use all the information from our
  original series of 1000 returns and to
  obtain $N=255$ VaR estimates, roughly corresponding to one trading year.
  In table~\ref{tab:backtest} are reported the results for the LTO.MI series. Our VaR models
  perform reasonably well with respect to both Kupiec's and Christoffersen's tests;
  the only exception is represented by the $\boldsymbol{\sigma^2}$-PPM model with
  $\alpha=5\%$, which produced $n=5$ exceptions, fairly low with respect to the
  expected number $\uE{n}=255\times0.05\approx 13$.
  The reason of this pitfall has to be located in the behaviour of our returns series;
  actually, an empirical study of that series has shown that the associated high frequency volatility
  decreases almost monotonously with time.
  Since the algorithm is trained with those returns corresponding to the high volatility regime
  and is tested against returns in a low volatility regime, this consequently results in a quite
  conservative evaluation of VaR. A similar behaviour is noticed also for the other series.

  \section{Concluding Remarks and Future Research}\label{s:conclusions}

  In this paper we have presented a novel Bayesian methodology for
  VaR computation  based on parametric PPMs. The
  main advantages of our approach are that it allows us to remain in
  the Normal setting, to identify anomalous observations and to obtain a closed-form expression for the
  VaR measure. This expression generalizes the standard parametric
  formula that is used in the literature under the normality
  assumption.  By means of PPMs we induce a clustering structure
  over the vector of means ($\boldsymbol{\mu}$-PPM) and we find the
  best agreement with ML approaches for significance level of order
  5\%. For lower values of $\alpha$ we obtained the best result by
  applying the PPMs to the vector of variances
  ($\boldsymbol{\sigma^2}$-PPM).

  We are currently working on the extension of the $\boldsymbol{\sigma^2}$-PPM approach to the
  portfolio analysis. The increase in the number of assets translates into an augmented dimensionality
  of the problem. In fact, the vector of variances is now replaced by the vector of covariance matrices.
  In order to reduce the number of involved parameters we are exploring several filtering techniques,
  see e.g. \cite{Laloux:1999}, \cite{Plerou:1999}, and \cite{Tumminello:2007}.

  \section*{Acknowledgements}

  The authors are thankful to  Guido Montagna and Oreste Nicrosini for helpful comments and suggestions.
  Maria Elena De Giuli and Claudia Tarantola acknowledge the University of Pavia for partial support.

  \newpage

  \begin{table}[p]
    \caption{Daily estimated VaR $(\%)$ values at 5\% and 1\% significance level with 68\% credible intervals.}
    \begin{tabular}{l c c c c c c}
      \hline
      \hline
      \multicolumn{1}{l}{VaR(\%)} & \multicolumn{3}{c}{$\alpha$=5\%} & \multicolumn{3}{c}{$\alpha$=1\%}\\
      \cline{2 - 7}
      & $\boldsymbol{\mu}$-PPM & $\boldsymbol{\sigma^2}$-PPM & $\boldsymbol{\sigma^2}$-CP
      & $\boldsymbol{\mu}$-PPM & $\boldsymbol{\sigma^2}$-PPM & $\boldsymbol{\sigma^2}$-CP\\
      \hline
      MIB30.MI & $1.45_{-0.05}^{+0.05}$ & $1.74^{+0.11}_{-0.12}$ & $1.76_{-0.01}^{+0.01}$ &
      $2.07_{-0.06}^{+0.06}$ & $2.48^{+0.16}_{-0.17}$ & $2.49_{-0.01}^{+0.01}$\\
      LTO.MI & $2.08_{-0.07}^{+0.07}$ & $2.78^{+0.15}_{-0.16}$ & $2.66_{-0.02}^{+0.02}$ &
      $2.95_{-0.09}^{+0.09}$ & $3.94^{+0.21}_{-0.21}$ &$3.78_{-0.03}^{+0.03}$\\
      MB.MI & $1.91_{-0.08}^{+0.07}$ & $2.40^{+0.12}_{-0.12}$ & $2.36_{-0.01}^{+0.01}$ &
      $2.72_{-0.11}^{+0.11}$ & $3.40^{+0.17}_{-0.17}$ & $3.35_{-0.02}^{+0.02}$\\
      SRG.MI & $1.58_{-0.05}^{+0.05}$ & $1.97^{+0.12}_{-0.13}$ & $2.01_{-0.01}^{+0.01}$ &
      $2.26_{-0.06}^{+0.06}$ & $2.81^{+0.17}_{-0.17}$ & $2.87_{-0.02}^{+0.02}$\\
      \hline
    \end{tabular}\label{tab:VaR}
  \end{table}

  \newpage

  \begin{table}[p]
    \caption{Posterior mean of the number of clusters and relative weight of the largest
      cluster for $\boldsymbol{\mu}$-PPM and $\boldsymbol{\sigma^2}$-PPM.}
    \begin{tabular}{l c c c c}
      \hline
      \hline
      & \multicolumn{2}{c}{Number of Clusters} & \multicolumn{2}{c}{Largest Cluster Weight} \\
      \cline{2-5}
      & $\boldsymbol{\mu}$-PPM & $\boldsymbol{\sigma^2}$-PPM &
      $\boldsymbol{\mu}$-PPM & $\boldsymbol{\sigma^2}$-PPM\\
      MIB30.MI & 3.11 & 3.39 & 0.986 & 0.990 \\
      LTO.MI & 5.02 & 4.52 & 0.963 & 0.944 \\
      MB.MI & 4.11 & 3.72 & 0.968 & 0.970\\
      SRG.MI & 3.44 & 3.59 & 0.984 & 0.978 \\
      \hline
    \end{tabular}\label{tab:clusters}
  \end{table}

  \newpage

  \begin{table}[p]
    \caption{
      Daily ML estimated VaR(\%) values at 5\% and 1\% significance level with 68\% bootstrap intervals.
      In the last column we report central value and 68\% bootstrap interval for the tail index $\nu$ of the GST.
    }
    \begin{tabular}{l c c c c | c}
      \hline
      \hline
      \multicolumn{1}{l}{VaR(\%)} & \multicolumn{2}{c}{$\alpha$=5\%} & \multicolumn{2}{c}{$\alpha$=1\%} &\\
      \cline{2 - 6}
      & Normal & Student-$t$ & Normal & Student-$t$ & $\nu$\\
      \hline
      MIB30.MI & $1.38_{-0.07}^{+0.05}$ & $1.27_{-0.06}^{+0.04}$ & $1.95_{-0.09}^{+0.07}$ &
      $2.22_{-0.10}^{+0.09}$ & $4.16^{+0.43}_{-0.48}$ \\
      LTO.MI & $2.50_{-0.15}^{+0.14}$ & $2.15_{-0.09}^{+0.08}$ & $3.55_{-0.20}^{+0.19}$ &
      $4.05_{-0.22}^{+0.19}$ & $3.26^{+0.28}_{-0.30}$ \\
      MB.MI & $2.07_{-0.09}^{+0.06}$ & $1.89_{-0.07}^{+0.05}$ & $2.95_{-0.11}^{+0.09}$ &
      $3.37_{-0.14}^{+0.12}$ & $3.93^{+0.35}_{-0.38}$ \\
      SRG.MI & $1.62_{-0.08}^{+0.05}$ & $1.48_{-0.07}^{+0.04}$ & $2.32_{-0.10}^{+0.08}$ &
      $2.65_{-0.12}^{+0.11}$ & $3.97^{+0.44}_{-0.44}$ \\
      \hline
    \end{tabular}\label{tab:MLVaR}
  \end{table}

  \newpage
  \begin{table}[p]
    \caption{Estimated 10-day VaR(\%) values at 5\% and 1\% significance level with 68\% credible intervals.}
    \begin{tabular}{l c c c c}
      \hline
      \hline
      \multicolumn{1}{l}{VaR(\%)} & \multicolumn{2}{c}{$\alpha$=5\%} & \multicolumn{2}{c}{$\alpha$=1\%}\\
      \cline{2 - 5}
      & $\boldsymbol{\mu}$-PPM & $\boldsymbol{\sigma^2}$-PPM
      & $\boldsymbol{\mu}$-PPM & $\boldsymbol{\sigma^2}$-PPM\\
      \hline
      MIB30.MI & $4.19^{+0.41}_{-0.40}$ & $4.43^{+0.46}_{-0.46}$ & $5.98^{+0.51}_{-0.51}$ &
      $6.33^{+0.59}_{-0.61}$\\
      LTO.MI & $6.35^{+0.69}_{-0.70}$ & $7.59^{+0.83}_{-0.85}$ & $9.08^{+0.89}_{-0.90}$ &
      $10.90^{+1.11}_{-1.13}$\\
      MB.MI & $6.55^{+0.65}_{-0.65}$ & $7.03^{+0.70}_{-0.71}$ & $9.41^{+0.82}_{-0.80}$ &
      $10.07^{+0.92}_{-0.91}$\\
      SRG.MI & $4.49^{+0.47}_{-0.47}$ & $4.83^{+0.54}_{-0.55}$ & $6.60^{+0.59}_{-0.59}$ &
      $7.06^{+0.70}_{-0.72}$\\
      \hline
    \end{tabular}\label{tab:10MLVaR}
  \end{table}

  \newpage

  \newpage

  \begin{sidewaystable}[p]\caption{ Sensitivity analysis of  outliers detection with respect to the value of
     $c$. $a$, $\lambda_0$, $\nu_0$ assume the values quoted in the
     main text. Subscripts $_{1,2,3}$ mean the corresponding cluster being equal to $S_1$, $S_2$, and $S_3$
     respectively (see section~\ref{s:Out}).
     }\label{tab:outliers}
     \vspace{0.2cm}
     \begin{tabular}{l l l}
        \hline
        \hline
        \multicolumn{1}{l}{c} & \multicolumn{1}{c}{MIB30} & \multicolumn{1}{c}{SRG.MI}\\
        \hline
        0.1 & $\{(46,48,155,476)_1(37,275)_2\}$ & $\{(6,7,46,47,108,155,271,782)_1(3,138,164,633,827)_3\}$ \\
        0.5 & $\{(46,48,155,476)_1(37,275,479,761)_2\}$ &$\{(6,7,46,47,108,155,271,782)_1(3,138,164,633,827)_3\}$  \\
        1 & $\{(37,46,48,155,275,476)_1 (5,45)_3\}$ &$\{(6,7,46,47,108,155,271,782)_1(3,138,164,633,827)_3\}$  \\
        5 & $\{(37,46,48,155,275,476)_1(5,45)_3\}$ & $\{(6,7,46,47,108,155,271,782)_1(3,138,164,633,827)_3\}$  \\
        10 & $\{(37,46,48,155,275,476)_1(5,45)_3\}$ &$\{(6,7,46,47,108,155,271,782)_1(3,138,164,633,827)_3\}$  \\
        50 & $\{(37,46,48,155,275,476,479,761)_1(5,32,45)_3\}$ &$\{(6,7,46,47,108,155,271,782)_1(3,138,164,633,827)_3\}$  \\
        \hline
        \multicolumn{1}{l}{c} & \multicolumn{2}{c}{LTO.MI} \\
        \hline
        0.1 & \multicolumn{2}{l}{$\{(7,11,422,464,742,787)_1(2,51,131,227,326,461,551,553,664,679,733,777,794)_3\}$}\\
        0.5 & \multicolumn{2}{l}{$\{(7,11,422,464,742,787)_1(2,51,131,227,326,461,551,553,664,679,733,777,794)_3\}$}\\
        1 & \multicolumn{2}{l}{$\{(7,11,422,464,742,787)_1(2,51,131,227,326,461,551,553,664,679,724,733,777,794)_3\}$}\\
        5 & \multicolumn{2}{l}{$\{(7,11,422,464,742,787)_1(2,51,131,227,326,461,551,553,664,679,724,733,777,794)_3\}$}\\
        10 & \multicolumn{2}{l}{$\{(7,11,422,464,742,787)_1(2,51,131,227,326,461,551,553,664,679,724,733,777,794)_3\}$}\\
        50 & \multicolumn{2}{l}{$\{(7,11,110,134,155,181,189,291,422,464,742,787)_1$}\\
        & \multicolumn{2}{l}{~~$(2,51,131,227,326,461,551,553,664,679,724,
        733,777,794)_3\}$}\\
        \hline
        \multicolumn{1}{l}{c} & \multicolumn{2}{c}{MB.MI} \\
        \hline
        0.1 & \multicolumn{2}{l}{$\{(48,106,237,370,515,530,540,573)_2(5,8,161,403,667,668,711,722)_3\}$}\\
        0.5 & \multicolumn{2}{l}{$\{(9,49)_1(5,8,48,106,161,237,370,403,515,530,540,573,667,668,711,722)_3\}$}\\
        1   & \multicolumn{2}{l}{$\{(9,49)_1(5,8,48,106,161,237,370,403,515,530,540,573,667,668,711,722)_3\}$}\\
        5   & \multicolumn{2}{l}{$\{(9,49,721)_1(5,8,48,106,161,237,370,403,515,530,540,573,667,668,711,722,813)_3\}$}\\
        10  & \multicolumn{2}{l}{$\{(9,49,721)_1(5,8,48,106,161,237,370,403,515,530,540,573,667,668,711,722,813)_3\}$}\\
        50  & \multicolumn{2}{l}{$\{(9,49,721)_1(5,8,46,48,106,161,237,370,403,428,515,530,540,573,667,668,
        710,711,722,813)_3\}$}\\
        \hline
     \end{tabular}
  \end{sidewaystable}

 \newpage

  \begin{table}[p]
    \caption{Backtesting results: the model is rejected at $5\%$ significance level
    if $LR_{UC}> 3.84$ (unconditional coverage test), or $LR_{CC}>5.99$
    (conditional coverage test).}
    \begin{tabular}{l c c c c c c}
      \hline
      \hline
      \multicolumn{1}{l}{LTO.MI} & \multicolumn{3}{c}{$\alpha$=1\%} & \multicolumn{3}{c}{$\alpha$=5\%}\\
      \cline{2 - 7}
      & \# Exceptions & $LR_{UC}$ & $LR_{CC}$ & \# Exceptions & $LR_{UC}$ & $LR_{CC} $\\
      \hline
      $\boldsymbol{\mu}$-PPM  & 5 & 1.857 & 2.057 & 9 & 1.288 & 1.947 \\
      $\boldsymbol{\sigma^2}$-PPM  & 1 & 1.237 & 1.245 & 5 & 13.873 & 14.073 \\
      \hline
    \end{tabular}\label{tab:backtest}
  \end{table}


\begin{thebibliography}{}
  \bibitem[Antoniak (1974)]{Antoniak:1974} Antoniak, C.E., 1974. Mixtures of Dirichlet processes with applications
	  to Bayesian nonparametric problems. \emph{Ann. Statist.} {\bf 2}, 1152 - 1174.
  \bibitem[Barry and Hartigan (1992)]{Barry_Hartigan:1992} Barry, D., Hartigan, J.A., 1992. Product partition models
	  for change point problems. \emph{Ann. Statist.} {\bf 20}, 260 - 279.
  \bibitem[Basel Committee (2006)]{Basel:2006} Basel Committee on Banking Supervision, 2006.
          \emph{Basel II: International convergence of capital measurement and capital standards. A revised framework}.
  \bibitem[Bormetti et al. (2007)]{Bormetti:2007} Bormetti, G., Cisana. E., Montagna, G., Nicrosini, O., 2007.
	  A non-Gaussian approach to risk measures. \emph{Physica A} {\bf 376}, 532 - 542.
  \bibitem[Bush and MacEachern (1996)]{Bush_MacEachern:1996} Bush, C.A., MacEachern, S.N., 1996. A semiparametric
	  Bayesian model for randomised block designs. \emph{Biometrika} {\bf 83}, 275 - 285.
  \bibitem[Chang and Feigenbaum (2008)]{Chang_etal:2008} Chang, G., Feigenbaum, J., 2008.
	  Detecting log-periodicity in a regime-switching model of stock returns. \emph{Quant. Finance} {\bf 8}, 723 - 738.
  \bibitem[Christoffersen (1998)]{Christoffersen:1998} Christoffersen, P.F., 1998. Evaluating interval forecasts.
	  \emph{Int. Econ. Rev.} {\bf 39}, 841-862.
  \bibitem[Dan\'ielsson et al. (1998)]{Danielsson:1998} Dan\'ielsson, J., Hartmann, P., de Vries, C.G., 1998. The cost
	  of conservatism: Extreme returns, Value-at-Risk, and the basle ``multiplication factor''. \emph{Risk} January.
  \bibitem[De Giuli et al. (2009)]{DeGiuli_etal:2009} De Giuli, M.E., Maggi, M.A., Tarantola, C., 2009.
	  Bayesian outlier detection in Capital Asset Pricing Model. \emph{Stat. Model.}, to appear.
  \bibitem[Gelfand and Smith (1990)]{Gelfand:1990} Gelfand, A.E., Smith, F.M., 1990.
	  Sampling-based approaches to calculating marginal densities. \emph{J. Amer. Statist. Assoc.} {\bf 85}, 398 - 409.
  \bibitem[Geman and Geman (1984)]{Geman:1984} Geman, S., Geman, D., 1984. Stochastic relaxation, Gibbs distributions, and Bayesian
	  restoration of images. \emph{IEEE Trans. Pattn Anal. Mach. Intell.} {\bf 6}, 721 - 741.
  \bibitem[Hartigan (1990)]{Hartigan:1990} Hartigan, J.A., 1990. Partition models. \emph{Commun. Statist. Theory Methods} {\bf 19}, 2745 - 2756.
  \bibitem[Hastings (1970)]{Hastings:1970} Hastings, W.K., 1970. Monte Carlo sampling methods using Markov chains and their applications.
	  \emph{Biometrika} {\bf 57}, 97 - 109.
  \bibitem[Jorion (2001)]{Jorion:2001} Jorion, P. 2001. \emph{Value at Risk: The new benchmark for managing financial risk.} McGraw-Hill: New York.
  \bibitem[Laloux et al. (1999)]{Laloux:1999} Laurent,  L., Cizeau, P., Bouchaud, J.P., Potters, M. 1999. Noise dressing of financial correlation matrices.
	  \emph{Phys. Rev. Lett.} {\bf 83}, 1467 - 1470.	  
  \bibitem[Kupiec (1995)]{Kupiec:1995} Kupiec, P.H., 1995. Techniques for verifying the accuracy of risk
	  measurement models. \emph{J. Derivatives} {\bf 3}, 73-84.
  \bibitem[Liu (1996)]{Liu:1996} Liu, J.S., 1996. Nonparametric hierarchical Bayes via sequential
	  inputations. \emph{Ann. Statist.} {\bf 24}, 911 - 930.
  \bibitem[Loschi et al. (2003)]{Loschi_etal:2003} Loschi, R.H., Cruz, F.R.B., Iglesias, P.L., Arellano-Valle, R.B.,
    2003. A Gibbs sampling scheme to the product partition model: An application to change-point problems.
    \emph{Comp. Oper. Res.} {\bf 30}, 463 - 482.
  \bibitem[Loschi et al. (2007)]{Loschi_etal:2007} Loschi, R.H., Iglesias, P.L., Arellano-Valle, R.B.,
    Cruz, F.R.B., 2007. Full predictivistic modeling of stock market data: Application of change point problems.
    \emph{Eur. J. Oper. Res.} {\bf 180}, 282 - 291.
  \bibitem[Loschi et al. (2008)]{Loschi_etal:2008} Loschi, R.H., Cruz, F.R.B., Takahashi, R.H.C.,
    Iglesias, P.L., Arellano-Valle, R.B., MacGregor Smith, J. , 2008. A note on Bayesian identification of change points in data sequences.
    \emph{Comp. Oper. Res.} {\bf 35}, 156 - 170.
  \bibitem[Manganelli and Engle (2004)]{Manganelli:2004} Manganelli, S., Engle, R.F. 2004. 
%	  A comparison of Value-at-Risk models in finance. 
          In \emph{Risk Measures for the 21st Century}, Szeg\"o G. Ed. Wiley: Chichester.
  \bibitem[Mattedi et al. (2004)]{Mattedi_etal:2004} Mattedi, A.P., Ramos, F.M., Rosa, R.R., Mantegna, R.N. , 2004.
	  Value-at-risk and Tsallis statistics: Risk analysis of the aerospace sector. \emph{Physica A} {\bf 344}, 554 - 561.
  \bibitem[Metropolis et al. (1953)]{Metropolis_etal:1953} Metropolis, N., Rosenbluth, A.W., Rosenbluth, M.N., Teller, A.H.,
	  Teller, E. 1953. Equations of state calculations by fast computing machines. \emph{J. Chem. Phys.} {\bf 21}, 1087 - 1091.
  \bibitem[Mina and Xiao (2001)]{Mina_Xiao:2001} Mina, J., Xiao, J.Y., 2001. \emph{Return to RiskMetrics. The
	  evolution of a standard.} RiskMetrics Group, New York.
  \bibitem[Plerou et al. (1999)]{Plerou:1999} Plerou, V., Gopikrishnan, P., Rosenow, B., Amaral, L.A.N., Stanley, H. E. (1999).
	  Universal and nonuniversal properties of cross correlations in financial time series. \emph{Phys. Rev. Lett.} {\bf 83}, 1471 - 1474.	  
  \bibitem[Quintana and Iglesias (2003)]{Quintana_Iglesias:2003} Quintana, F.A., Iglesias, P.L., 2003.
	  Bayesian clustering and product partition models. \emph{J. Roy. Statist. Soc. B} {\bf 65}, 557 - 574.
  \bibitem[Quintana et al. (2005)]{Quintana_etal:2005} Quintana, F.A., Iglesias, P.L., Bolfarine, H., 2005.
	  Bayesian identification of outliers and change-points in measurement error models. \emph{Adv. Complex Syst.} {\bf 8}, 433 - 449.
  \bibitem[Smith (2001)]{Smith:2001}  Smith, B. (2001) Bayesian Output Analysis program: version
    1.0.0. Dept. of Biostatistics. University of Iowa, USA.
    Available at \emph{http://www.public-health.uiowa.edu/boa}.
  \bibitem[Tarantola et al. (2008)]{Tarantola_etal:2008} Tarantola, C., Consonni, G., Dellaportas, P., 2008.
	  Bayesian clustering for row effects models. \emph{JSPI} {\bf 138}, 2223 - 2235.
  \bibitem[Tumminello et al. (2007)]{Tumminello:2007} Tumminello, M., Lillo, F., Mantegna, R.N. 2007.
	  Kullback-Leibler distance as a measure of the information filtered from multivariate data. \emph{Phys. Rev. E} {\bf 76}, 031123.	  
  \end{thebibliography}
\end{document}